\documentclass[english, 11pt]{article}
\usepackage[T1]{fontenc}
\usepackage[latin9]{inputenc}
\usepackage{geometry}
\usepackage{float}
\usepackage{graphicx}
\usepackage{babel}
\usepackage{amssymb}
\usepackage{amsmath}
\usepackage{subfig}
\usepackage{appendix}
\usepackage{color}
\usepackage{multirow}
\usepackage{longtable}
\usepackage{booktabs}
\usepackage{array}
\usepackage{tablefootnote}
\usepackage{longtable}
\usepackage{url}

\newcommand{\PreserveBackslash}[1]{\let\temp=\\#1\let\\=\temp}
\newcolumntype{C}[1]{>{\PreserveBackslash\centering}p{#1}}
\newcolumntype{R}[1]{>{\PreserveBackslash\raggedleft}p{#1}}
\newcolumntype{L}[1]{>{\PreserveBackslash\raggedright}p{#1}}

\newcommand{\beginsupplement}{%
        \setcounter{table}{0}
        \renewcommand{\thetable}{S\arabic{table}}%
        \setcounter{figure}{0}
        \renewcommand{\thefigure}{S\arabic{figure}}%
        \renewcommand{\thesection}{S}%
     }

\begin{document}

\title{Spatio-temporal quasi-experimental methods for rare disease outcomes: The impact of reformulated gasoline on childhood hematologic cancer}

\author{Sofia L. Vega and Rachel C. Nethery\\
Department of Biostatistics, Harvard T.H. Chan School of Public Health}
\date{}

\maketitle
\begin{abstract}

Although some pollutants emitted in vehicle exhaust, such as benzene, are known to cause leukemia in adults with high exposure levels, less is known about the relationship between traffic-related air pollution (TRAP) and childhood hematologic cancer. In the 1990s, the US EPA enacted the reformulated gasoline program in select areas of the US, which drastically reduced ambient TRAP in affected areas. This created an ideal quasi-experiment to study the effects of TRAP on childhood hematologic cancers. However, existing methods for quasi-experimental analyses can perform poorly when outcomes are rare and unstable, as with childhood cancer incidence. We develop Bayesian spatio-temporal matrix completion methods to conduct causal inference in quasi-experimental settings with rare outcomes. Selective information sharing across space and time enables stable estimation, and the Bayesian approach facilitates uncertainty quantification. We evaluate the methods through simulations and apply them to estimate the causal effects of TRAP on childhood leukemia and lymphoma. 

\vspace{.2cm}

\end{abstract}

\vspace{.3in}
\section{Introduction}

Traffic-related air pollution (TRAP) has long been a suspected risk factor for childhood hematologic cancers. Pollutants emitted in vehicle exhaust, such as benzene and 1,3-butadiene, are known to cause leukemia in adults with high occupational exposure levels \cite{rinsky1981leukemia,rinsky1987benzene}, but meta-analyses and reviews of the large literature studying the relationship between ambient TRAP and pediatric hematologic cancer have failed to establish a causal relationship \cite{raaschou2006air,boothe2014residential,sun2014no,filippini2015review}. The inconclusive results are likely due to the rarity of childhood cancer, the difficulty in measuring ambient TRAP exposure and exposure timing, and confounding \cite{buffler2005environmental,linet2005etiology}. 

In the 1990s, in an effort to control TRAP, the US EPA enacted gasoline content regulations in select heavily-polluted areas of the US. In particular, the regulations mandated that only reformulated gasoline be sold in these areas. In reformulated gasoline, the allowable benzene content was capped at 1\% by volume, and it was required to contain at least 2\% oxygen via the use of an oxygenate (such as ethanol) \cite{auffhammer2011clearing}. The reformulated gasoline was intended to reduce volatile organic compound and toxic air pollutant emissions by 15\% when compared to conventional gasoline, without increasing NO$_x$ emissions  \cite{auffhammer2011clearing}. This new limit on mobile-source benzene induced substantial declines in ambient TRAP in affected areas. For example, California began selling reformulated gasoline between 1995-1996. It is estimated that ambient benzene concentrations decreased by 30\%-40\% between 1994-1996 as a result \cite{propper2015ambient, harley2006effects}. These reformulated gasoline regulations created an ideal quasi-experiment (QE) to study the effects of TRAP on childhood hematologic cancer incidence, as regulated areas experienced sudden changes in exposure while many nearby areas did not. Quasi-experimental studies are advantageous in environmental health contexts because they enable use of inexpensive panel data and leverage localized exposure changes to estimate health effects adjusted for certain types of unmeasured confounding. In this paper, our aim is to develop analytic and inferential methods for quasi-experimental studies with rare disease incidence outcomes and apply these methods to study the impact of the reformulated gasoline program on childhood hematologic cancer incidence.

A wide range of approaches, relying on different causal identifying assumptions, have been proposed for estimating effects in QE settings. Difference-in-differences is a widely-used approach that relies on the parallel trends assumption, which states the difference in outcomes in the treated and control groups would have remained constant over the study period in the absence of treatment. Parallel trends can also be described as the assumption of no time-varying unmeasured confounders. Under this assumption, causal effects are estimated by comparing the observed outcome trend in the control group to the observed outcome trend in the treatment group. However, in the presence of time-varying confounders, the parallel trends assumption is violated. This limitation gave rise to the synthetic control method \cite{abadie2010synthetic}, which controls for certain types of time-varying confounders by leveraging pre-treatment temporal trends in treated and control units. It achieves this by reweighting control units to construct a ``synthetic control unit'' whose post-treatment outcomes serve as the counterfactual outcomes for a single treated unit. The weights are constructed to ensure that pre-treatment trends in the synthetic control unit mimic those in the treated unit, which accounts for trends induced by certain time-varying confounders.

In recent years, various methods have been introduced to improve or extend difference-in-differences and synthetic control methods. Doubly-robust approaches to effect estimation in QE settings have been proposed \cite{arkhangelsky2021double,arkhangelsky2021synthetic}, which utilize two different modeling strategies to adjust for confounding bias, and are able to recover causal effects if either model is correctly specified. 
The Augmented Synthetic Control Method \cite{Ben2021} adds a bias-correction to the standard synthetic control method to improve causal effect estimates when the synthetic control unit does not adequately capture pre-treatment trends in the treated unit.

An adjacent literature has emerged proposing causal inference methods for QEs with panel data that model outcomes explicitly, then employ these models to impute missing counterfactual outcomes. In this paper, we build on the matrix completion (MC) approach for causal inference with panel data, proposed by Athey et al \cite{athey2021matrix} (a closely related approach called generalized synthetic controls was introduced by Xu et al. \cite{xu2017generalized}). Like synthetic controls, this method leverages temporal trends in treated and control units to adjust for time-varying unmeasured confounders, but it allows for multiple treatment units and variable treatment initiation times. The approach proceeds as follows: (1) structure the panel data into a matrix; (2) set outcomes for treated units during post-treatment period to missing to create a matrix of outcomes under control; (3) apply MC to this matrix to obtain a low-rank matrix factorization; (4) use the matrix factorization to impute the missing outcomes in the matrix. The imputed outcomes represent estimates of the counterfactuals, i.e., the outcomes expected in the treated units at post-treatment times, in the absence of treatment. MC can adjust for time-varying unmeasured confounders, so long as their temporal trends do not change differentially in treated and control units at post-treatment times.

MC and synthetic control methods rely heavily on capturing pre-treatment trends in outcomes in order to adjust for confounding. The performance of these methods has not been evaluated in the context of rare disease outcome data, where trends are unstable and noisy. However, because they rely on either stable patterns across time or stable patterns across units ${\cite{athey2021matrix}}$, they are unlikely to be able to properly adjust for confounding when outcomes are rare and trends are highly impacted by noise.

Another limitation of standard MC and synthetic control methods is the lack of reliable, finite sample inferential tools. In synthetic control methods, researchers can, in some cases, perform inference using a permutation test  \cite{RePEc:gam:jecnmx:v:5:y:2017:i:4:p:52-:d:120610}.  However, these tests often fail, as the requisite symmetry assumption for permutation tests is commonly violated \cite{RePEc:gam:jecnmx:v:5:y:2017:i:4:p:52-:d:120610}. Recently, Chernozhukov et al. \cite{chernozhukov2021exact} make use of ideas from conformal prediction and permutation tests for an inference procedure that can be generalized for many different panel data methods. Although literature has emerged expanding on MC for causal inference using non-Bayesian estimation approaches \cite{farias2021learning, choi2023matrix, agarwal2023causal, agarwal2020synthetic, agarwal2022network}, very limited recent literature has begun to explore Bayesian implementations of MC for panel data \cite{pang2021bayesian,tanaka2021bayesian}. The Bayesian approach is advantageous relative to frequentist implementations, as inference does not require strict assumptions and can be provided in a consistent and straightforward procedure through the posterior predictive distribution of treated units' counterfactuals \cite{tanaka2021bayesian,pang2021bayesian,nethery2021integrated}. 

Integrating concepts from the Bayesian disease mapping and latent factor model literatures, we seek to extend MC methods to enable robust investigation of the impacts of environmental QEs on rare disease outcomes. Our main contributions are (1) evaluating the performance of existing MC and synthetic control methods in rare outcome contexts and (2) proposing alternative Bayesian MC approaches using spatio-temporal smoothing, regularization, and adaptive hyperparameter selection to reduce over-fitting and improve estimation in this setting. Through an extensive simulation study designed to mimic real disease outcome scenarios, we assess the performance of our proposed methods and compare them to existing MC and synthetic control methods. We then apply these methods to study the effects of TRAP on the incidence of childhood and young adult leukemia and lymphoma using the reformulated gasoline program QE described above. Although our methods are directly motivated by this case study, they can be used more generally when trends in outcomes are unstable/noisy or when spatial or temporal correlation is expected. 

\section{Methods}
 
\subsection{Motivating Data}

The federal reformulated gasoline program targeted areas with severe ozone problems, mandating the sale of exclusively reformulated gasoline in certain metropolitan counties within the following states: California, Connecticut, New Jersey, New York, Delaware, Maryland, Pennsylvania, Indiana, Texas, and Wisconsin. The full list of counties included in the program is shown in Supplemental Table \ref{tab:FedRFG} \cite{epa}. The program took effect in 1995 for the counties listed in this table. The state of California also implemented its own reformulated gasoline program, requiring that gasoline with similar requirements to the federal reformulated gasoline be sold statewide \cite{national1999ozone}, which went into effect in the summer of 1996.

In general, in our analyses, we consider a county to be ``treated'' if and when the reformulated gasoline program went into effect there. However, because some counties in California began using reformulated gasoline in 1995 and others in 1996, there is concern of spillover effects during the interim period. It is possible the effect of using the reformulated gas spilled over to counties that hadn't enacted the regulation (or vice-versa) through commuting and migration. Additionally, there is concern of anticipation. It is possible some gas stations in untreated California counties in 1995 began selling reformulated gasoline in anticipation of the program implementations. For this reason, in our primary analyses we consider the entire state of California to have been treated starting in 1995.

The study period of interest is 1988 to 2003, to incorporate a substantial number of years pre- and post-treatment to discern trends in the outcome. We rely primarily on childhood and young adult (CYA) leukemia and lymphoma incidence data from the National Cancer Institute's Surveillance Epidemiology End Result (SEER) database \cite{seer}. During the study period, 9 unique cancer registries contributed population-level data to SEER, covering the following areas: San Francisco Bay Area (California), Connecticut, Atlanta (Georgia), Hawaii, Iowa, Detroit (Michigan), New Mexico, Utah, and Seattle-Puget Sound (Washington). Note that the SEER-covered areas include counties that were both treated and untreated in the reformulated gasoline program.

Due to differing data availability by disease type, the analyses for leukemia and lymphoma cover slightly different study areas. For lymphoma, only SEER data are available, so the study area is the SEER-covered areas listed above. The treated counties in the lymphoma analyses thus consist of 5 California counties and 8 Connecticut counties. The control counties are the remaining 155 counties SEER-covered counties in Georgia, Hawaii, Idaho, Minnesota, New Mexico, Utah, and Washington. For leukemia, in addition to SEER, we have access to incidence data from the California Cancer Registry during the study period, which covers all California counties. Thus, we make use of both data sources in the leukemia analyses, and the set of treated counties is expanded in the leukemia analyses to include all 52 California counties. For computation purposes, we restrict our analyses to counties that have more than 1 case of leukemia or lymphoma in the pre-treatment period. This resulted in 131 counties for the leukemia analyses and 184 counties for the lymphoma analyses.

From these data sources, annual county-level leukemia and lymphoma incidence rates were constructed. Childhood leukemia is most commonly diagnosed in children ages 1-4, and over 60\% of new cases occur in children $<10$ years old \cite{seer}. Moreover, leukemia cases occurring in younger and older children tend to be biologically distinct \cite{dores2012acute,yamamoto2008patterns}. For these reasons, we study leukemia incidence among children ages 0-9, exclusively. The epidemiology of lymphoma in children and young adults is quite different from leukemia, with risk increasing monotonically up to the mid-to-late 20s \cite{uscs2022}. Thus, for our lymphoma analyses, we investigate incidence among 0-29 year olds.

Although MC can, in principle, adjust for time-varying unmeasured confounders that do not change differentially in treated and control units post-treatment, we explicitly control for one time-varying potential confounder in our analyses that may violate this condition: county-level percent Hispanic children and young adults. Hispanic children in the US are known to be at higher risk for hematologic malignancies \cite{whitehead2016childhood}, and the size of the Hispanic population is likely to have been changing differentially in treated and control areas during our study period. Thus, to avoid possible residual confounding from this variable after applying MC, we explicitly adjust for it in our analyses using data retrieved from the U.S. Census Bureau.

\subsection{Notation and Matrix Completion}

Consider our CYA cancer incidence panel data. Let $i=1,...,N$ index the counties of interest where $N$ is the number of counties and $t=1,...,T$ index the time periods in years where $T$ is the last year of the study. Suppose the outcome data are arranged into an $N \times T$ panel matrix, with counties in the rows and years in the columns. The observed outcome for county $i$ at time $t$ can be denoted as $Y_{it}$. Let $\mathcal{W}$ denote the set of indices for treated units and time periods and let $N_{\mathcal{W}}$ denote the number of treated units. 

We situate our approaches within the potential outcomes framework \cite{Rubin1974}, such that $Y_{it}(1)$ and $Y_{it}(0)$ denote the outcomes that would have been observed in county $i$ at time $t$ under treatment and control, respectively. For $(i,t) \in \mathcal{W}$, $Y_{it}$=$Y_{it}(1)$ , i.e., the observed outcome is equal to the potential outcome under treatment for treated units at post-treatment times (Figure 1). Conversely, for $(i,t) \notin \mathcal{W}$, $Y_{it}$=$Y_{it}(0)$. Because $Y_{it}(0)$ is unobserved for the treated units at post-treatment times, the aim of the methods introduced in this paper is to estimate $Y_{it}(0)$ for these units (which are then used to estimate treatment effects, i.e., $Y_{it}(1)-\widehat{Y}_{it}(0)=Y_{it}-\widehat{Y}_{it}(0)$). For the models used throughout this paper, we follow the causal identifying assumptions for Bayesian matrix completion as described by Tanaka et al., Nethery et al., and Pang et al. \cite{tanaka2021bayesian,nethery2021integrated,pang2021bayesian}. That is, we assume the stable unit treatment value assumption (SUTVA) \cite{rubin1980randomization}, causal consistency, latent ignorability, the unobserved time-varying confounders can be captured by a small number of latent factors, and conditional exchangeability. See \ref{sec:assumptions} for a more detailed description of these assumptions.

In the formulation of MC methods for causal inference with QEs, the matrix with missing values is the matrix of $Y_{it}(0)$, denoted $\mathbf{Y(0)}$. This is simply the panel data matrix with entries in the positions corresponding to the treated units at post-treatment times artificially set to missing. In Athey et al. \cite{athey2021matrix}, the missing post-treatment $Y_{it}(0)$ values are imputed using the factors obtained from a low-rank factorization of the matrix $\mathbf{Y(0)}$ into a matrix of time-specific latent factors $\mathbf{V}$ and unit-specific factor loadings $\mathbf{U}$. The MC model takes the form $Y_{it}(0)=\mathbf{U}_i^T\mathbf{V}_t+\epsilon_{it}$. Here, $\mathbf{U}_i$ is a $K$-length vector of county-specific latent factors, $\mathbf{V}_t$ is a $K$-length vector of unknown time-specific factor loadings, and $K$ is an unknown scalar with $K\leq \leq \mathrm{min}(N,T)$. The matrices $\mathbf{U} \in \mathbb{R}^{K \times N}$ and $\mathbf{V} \in \mathbb{R}^{K \times T}$ aim to capture unmeasured confounders in the data.

To implement MC in a Bayesian framework, one can use a likelihood of the form:
\begin{equation}
    P(\mathbf{Y(0)} \mid \mathbf{U},\mathbf{V},\Sigma) = \prod_i\prod_t[f(Y_{it}(0)|\mathbf{U}_i^T\mathbf{V}_t,\Sigma)]\times \mathrm{I}[(i, t) \notin \mathcal{W}]
\end{equation}
where $f$ is the pmf or pdf of the response with mean given by $\mathbf{U}_i^T\mathbf{V}_t$ and common scale parameter $\Sigma$.
Often, Gaussian priors with means $\mu_U$ and $\mu_V$ and precision matrices $\Lambda^{-1}_U$ and $\Lambda^{-1}_V$ are placed on $\mathbf{U}$ and $\mathbf{V}$, respectively \cite{salakhutdinov2008bayesian}.

The posterior predictive distribution for the missing values, $\mathbf{Y(0)}^{mis}$, is obtained by conditioning on the observed values, $\mathbf{Y(0)}^{obs}$, and marginalizing over the model parameters and hyperparameters:
$$
    P(\mathbf{Y(0)}^{mis}\mid \mathbf{Y(0)}^{obs}
    ) = \int P(\mathbf{Y(0)}^{mis}\mid \mathbf{U},\mathbf{V},\Sigma) \ P(\mathbf{U},\mathbf{V},\Sigma \mid \mathbf{Y(0)}^{obs}) \ d(\mathbf{U},\mathbf{V},\Sigma)
$$
Exact evaluation of this predictive distribution is analytically intractable. Instead, Markov Chain Monte Carlo (MCMC) based methods can use a Monte Carlo approximation to the predictive distribution by: 

$$
    P(\mathbf{Y(0)}^{mis}\mid \mathbf{Y(0)}^{obs}) \approx \frac{1}{M} \sum^M_{m=1}P(\mathbf{Y(0)}^{mis}\mid \mathbf{U}^{(m)},\mathbf{V}^{(m)},\Sigma) 
$$
where the samples of $\{\mathbf{U}^{(m)},\mathbf{V}^{(m)}\}$ are generated by running a Markov Chain whose stationary distribution is the posterior distribution over the model parameters and hyperparameters \cite{salakhutdinov2008bayesian}. Here, $\mathbf{U}$ and $\mathbf{V}$ are not identifiable without further constraints \cite{tanaka2021bayesian}. However, in the Bayesian context, this does not compromise the identifiability of the predictions from the model \cite{nethery2021integrated}.

\subsection{Matrix Completion Approaches for Rare Count Outcomes}

\begin{figure}
\begin{minipage}[]{1\columnwidth}
\begin{center}
\includegraphics[scale=0.115]{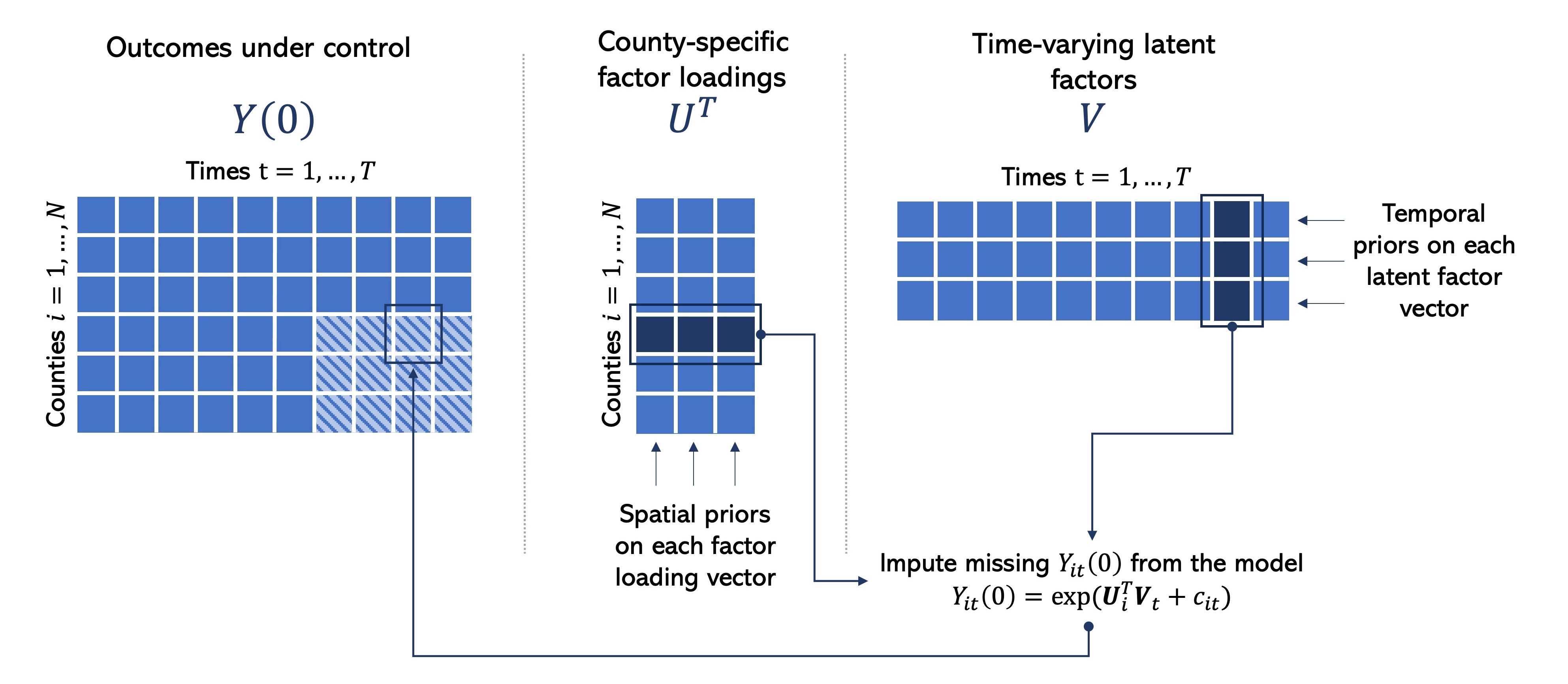}
\par\end{center}%
\end{minipage}\protect\caption{Illustration of spatio-temporal matrix completion. $c_{it}$ denotes all other components of the mean model.}
\label{fig:MC}
\end{figure}

Our model builds on the Bayesian MC models for Poisson/count data \cite{cemgil2009bayesian,gopalan2014content,liang2014codebook,nethery2021integrated} by tailoring it to QE analysis settings with rare outcomes, which are common in small-area cancer incidence data, and to accommodate complex spatio-temporal correlation structures in the outcomes. First, by taking a Bayesian approach, our models inherently regularize, which may help prevent over-fitting noisy data. Second, to account for the instability in trends over space and time that often occur with rare outcome data and may compromise MC's ability to adjust for unmeasured confounders, we introduce spatio-temporal priors into relevant components of the MC model (Figure \ref{fig:MC}).  The addition of these priors will help stabilize trends by sharing information across space and time. In the context of environmental exposures and disease outcomes, any unmeasured confounders (represented by the $\mathbf{U}$ and $\mathbf{V}$ matrices) would likely have a spatio-temporal structure. By explicitly encoding this structure into the $\mathbf{U}$ and $\mathbf{V}$ matrices, the model  is more likely to uncover these confounders. In some of our models, we also include a shrinkage prior that allows for adaptive selection of $K$, which reduces the chances of model misspecification.

The various model formulations we will propose leverage differing prior distributions but all impose the following mean structure for the $Y_{it}(0)$, which is similar to that of \cite{nethery2021integrated}:
\begin{equation}
    \mathrm{log}\left(E\left[Y_{it}(0)\right]\right)=\alpha + \gamma_i + \psi_t + \mathbf{U}_i^{T}\mathbf{V}_t +\mathbf{X}_{it}'\boldsymbol{\beta}+\mathrm{log}(\theta_{\mathrm{it}})
\end{equation} 
where $\alpha$ is a global intercept, $\gamma_i$ are county-specific deviations from the global intercept, $\psi_t$ are time-specific deviations from the global intercept,  $\mathbf{U}_i$ and $\mathbf{V}_t$ are defined as above, $\mathbf{X}_{it}$ is a vector of time-varying measured confounders (if any) and $\boldsymbol{\beta}$ is a vector of their coefficient parameters, and $\theta_{it}$ is a scalar offset of the population size at county $i$ and time point $t$. This model extends classic Bayesian MC methods by allowing for a non-identity link and an offset, all of which are critical when modeling and predicting the incidence of rare diseases.  

Since we are using count data in our application, our likelihood will take the form of equation (1) where $f$ is the negative binomial pmf with mean given by equation (2) and common scale parameter $\Sigma$. The missing entries in $\mathbf{Y(0)}$ are then sampled from the posterior predictive distribution. We then construct MCMC samples of the ATT for each treated time as:
$$
\mathrm{ATT}_t^{(m)} = \frac{1}{N_\mathcal{W}}\sum_{i \in W}\Big[ Y_{it}(1) - Y_{it}^{(m)}(0)\Big]
$$
where $Y_{it}^{(m)}(0)$ is the $m^{\mathrm{th}}$ MCMC sample of $Y_{it}(0)$. Using summaries of the MCMC samples, we can estimate $\mathrm{ATT}_t$ and its uncertainties. Each of our Bayesian models is implemented in the \texttt{rstan} R package ${\cite{rstan}}$, which uses Hamiltonian Monte Carlo to collect posterior samples. In Section \ref{sec:computation}, we discuss several computational considerations-- most notably issues that may arise when using \texttt{rstan} for spatial modeling, which have been noted in prior literature \cite{morris2019bayesian}, and the recommended workarounds.

With panel data, the success of MC models depends on the ability of the model to capture trends over space and time. Taking this into account, prior distributions on the latent factors and factor loadings matrices that account for spatio-temporal correlation will be utilized. That is, we consider spatial priors for the county-specific factor loadings in the rows of $\mathbf{U}$ ($\mathbf{U}_k$, $k=1,...,K$) and temporal priors on the time-specific latent factors in the rows of $\mathbf{V}$ ($\mathbf{V}_k$). Additionally, non-informative priors are specified for $\alpha$, $\gamma_i$, $\psi_t$, and $\beta$ i.e. $\Sigma \sim \mathrm{Unif}(0,\infty)$ and $\alpha, \gamma_i, \psi_t, \beta \sim \mathrm{Unif}(-\infty,\infty)$. A $\mathrm{N}(0,1)$ prior is placed on $\mathrm{log}(\Sigma)$.

Our methodological contribution centers on addressing the significant instability present in rare disease incidence data by incorporating spatial and temporal priors. Through the utilization of spatial and temporal smoothness constraints on the latent factor structure, we seek to alleviate the difficulties arising from extracting signal from highly variable time series. These constraints serve to direct the model in capturing smoothly evolving factors that are believed, based on substantive knowledge, to be most likely to represent latent variables driving trends in chronic disease.

In spatio-temporal models for areal data, spatially structured parameters and/or latent variables are typically obtained using variations of conditional autoregressive priors. Here, to allow for spatial correlation in parameters, we consider the widely-used intrinsic conditional auto-regressive (ICAR) prior, which assumes that the parameter corresponding to a given spatial unit is correlated with the parameters of its neighboring spatial units \cite{besag1995conditional}. Temporally structured parameters are often obtained by specifying auto-regressive priors. In this paper, we consider the first-order auto-regressive prior (AR(1))${\cite{engle1982autoregressive}}$ that accounts for temporal correlation in parameters by using the parameter value from the previous time point to smooth the current time point. We also consider a temporal ICAR prior that assumes the parameter for a given time point is correlated with parameters for both of its neighboring times. See Anderson et al. \cite{anderson2017comparison} for an in-depth review of spatio-temporal modeling.

To see how the choice of priors on the latent factors and factor loadings matrices impacts the ability of the model to recover the missing counterfactuals in rare outcome settings, we form six candidate models, characterized by different prior specifications on $\mathbf{U}_k$ and $\mathbf{V}_k$. We describe each of these models in detail below, and they are also summarized in Table~\ref{tab:models}. 

\begin{table}[H]
\centering
\caption{The six Bayesian MC models considered, as defined by different prior specifications on the factors and factor loadings.}
\label{tab:models}
\begin{tabular}[c]{ p{4cm} c p{3.5cm} p{5cm}  }
 \toprule
 \textbf{Model Name} & \textbf{Model \#} &  \textbf{Priors on }$\mathbf{U_k}$& \textbf{Priors on }$\mathbf{V_k}$\\
 \toprule
 Vanilla & 1  & Normal(0,1)    &Normal(0,1)\\
 
 Spatial & 2 &   Spatial ICAR  & Normal(0,1)\\

 Space-Time ICAR & 3 & Spatial ICAR & Temporal ICAR\\

 Space-Time AR(1) & 4 & Spatial ICAR&  Temporal AR(1)\\

 Space-Time Lasso & 5 & Laplace & Laplace\\

 Space-Time Shrinkage& 6 &   Spatial ICAR  & Temporal AR(1)-Shrinkage\\

 \toprule
\end{tabular}
\end{table}

\subsubsection{Model 1: The Vanilla Model}

The Vanilla Model does not formally account for spatial or temporal correlation, placing the conventional independent normal priors on the elements of $\textbf{U}_k$ and $\textbf{V}_k$, i.e., 
$\mathbf{U}_k \sim \mathrm{N}(\mathbf{0},\mathbf{I}_N)$ and 
$\mathbf{V}_k \sim \mathrm{N}(\mathbf{0},\mathbf{I}_T)$, where $\mathbf{I}_N$ and $\mathbf{I}_T$ are $N \times N$ and $T \times T$ identity matrices, respectively.

\subsubsection{Model 2: The Spatial Model}
The Spatial Model accounts for spatial correlation through a spatial prior placed on $\textbf{U}_k$, but does not account for temporal correlation. In this model, an Intrinsic Conditional Auto-Regressive (ICAR)${\cite{besag1995conditional}}$ prior that assumes complete spatial correlation between units is placed on $\textbf{U}_k$: 
$$
\mathbf{U}_{k} \sim \mathrm{N}(0,[\tau(D_S-W_S)]^{-1})
$$
where $W_S$ is the $N \times N$ spatial adjacency matrix where entries $\{i, i\}$ are zero and the off-diagonal elements are 1 if regions $i$ and $j$ are neighbors and 0 otherwise, and $D_S$ is the $N \times N$ diagonal matrix where entries $\{i, i\}$ are the number of neighbors of area $i$ and the off-diagonal entries are 0. We place a $\mathrm{Ga(1,0.01)}$ prior on $\tau^2$ following the default prior specification for this parameter in the \texttt{CARBayesST} R package \cite{JSSv084i09}. As in the Vanilla Model, standard normal priors are placed on the $\textbf{V}_k$.

\subsubsection{Model 3: Space-Time ICAR Model}
The Space-Time ICAR Model places ICAR priors on the rows of $\textbf{U}$ and $\textbf{V}$ to account for spatial and temporal correlation:
$$
\begin{gathered}
\mathbf{U}_{k} \sim \mathrm{N}(0,[\tau_S(D_S-W_S)]^{-1})\\
\mathbf{V}_{k} \sim \mathrm{N}(0,[\tau_T(D_T-W_T)]^{-1})
\end{gathered}
$$
Here the spatial ICAR prior is as defined in Model 2. In the temporal ICAR prior, $W_T$ is the $T \times T$ adjacency matrix where entries $\{i, i\}$ are zero and the off-diagonal elements are 1 if time-point $i$ and $j$ are neighbors and 0 otherwise, and $D_T$ is the $T \times T$ diagonal matrix where entries $\{i, i\}$ are the number of neighbors of time-point $i$ and the off-diagonal entries are 0.  Additionally, we place $\mathrm{Ga(1,0.01)}$ priors on $\tau_S$ and $\tau_T$.  

\subsubsection{Model 4: Space-Time AR Model}
The Space-Time AR Model accounts for spatial correlation through the placement of an ICAR prior on the rows of  $\textbf{U}$. To account for temporal correlation in $\mathbf{V}_{k}$, instead of a temporal ICAR prior we specify a First-Order Auto-Regressive prior (AR(1))${\cite{engle1982autoregressive}}$ that takes each point $V_{kt}$ in the rows of $\mathbf{V}$ to be generated according to 

$$
\begin{gathered}
    \mathbf{U}_{k} \sim \mathrm{N}(0,[\tau(D_S-W_S)]^{-1})\\
    V_{kt} \sim \mathrm{N}(a + b *V_{k(t-1)},\sigma) \\
\end{gathered}
$$
Here, improper flat priors (i.e. $\mathrm{Uniform}(-\infty,\infty)$) are placed on the coefficients, $a$ and $b$, and $\sigma$ is a positively constrained noise scale parameter with a $\mathrm{Uniform}(0,\infty)$ prior.

\subsubsection{Model 5: Space-Time Lasso Model}

The Space-Time Lasso Model induces sparsity by placing independent Laplace distributions as the prior distributions for the rows $\textbf{U}$ and $\textbf{V}$ \cite{masuda2022point}. These priors have previously been applied to detect clusters in spatial data by selecting groups of points that have similar values for covariates and are adjacent \cite{masuda2022point}. In our model, the priors encourage the fusion of the loadings and factors for units that are close in space and time, respectively. If we let $C_S$ denote a set of all adjacent pairs of spatial units and $C_T$ denote a set of all adjacent pairs of temporal units:

$$
\begin{gathered}
\mathbf{U}_{k} | \lambda_{U1},\lambda_{U2} \sim \prod_{(j,j')\in C_S} \mathrm{exp}[-\lambda_{U1}|U_j-U_{j'}|] * \prod^N_{i=1}\mathrm{exp}[-\lambda_{U2}|U_i|]\\
\mathbf{V}_{k} | \lambda_{V1},\lambda_{V2} \sim \prod_{(j,j') \in C_T} \mathrm{exp}[-\lambda_{V1}|V_j-V_{j'}|] * \prod^T_{t=1}\mathrm{exp}[-\lambda_{V2}|V_t|]\\
\end{gathered}
$$

Following Casella et al. \cite{casella2010penalized}, we place $\mathrm{Ga}(1,0.01)$ priors on $\lambda_{U1},\lambda_{U2},\lambda_{V1},$ and $\lambda_{V2}$ and enter them into the Gibbs sampler, as this method is fast and leads to posterior means close to the marginal maximum likelihood estimates.

\subsubsection{Model 6: Space-Time Shrinkage Model}
The Space-Time Shrinkage Model accounts for spatial correlation through the placement of an ICAR prior on the rows of $\textbf{U}$, as in Models 2 and 3 above. 
In standard factor models, the need for user-specification of the number of factors, $K$, poses challenges and introduces subjectivity. By specifying a $K$ too large, we risk over-fitting the data and add computational burden, and by specifying a $K$ too small, we have the chance to under-fit and waste information by leaving out important factors. To address this issue and eliminate the need for a user-specified $K$, Bhattacharya and Dunson ${\cite{bhattacharya2011sparse}}$ propose the multiplicative gamma shrinkage prior for latent factor models, which in theory allows for the introduction of infinitely many factors, with the factors increasingly shrunk towards 0 as the row index increases. This is achieved by introducing a scaling factor into the prior variance of the factors that converges towards zero with increasing $K$ at a rate informed by the data. This adaptive selection of $K$ reduces the chances of model misspecification and over-fitting. Here we unify the multiplicative gamma shrinkage prior with the AR(1) prior, to simultaneously account for temporal correlation in $\mathbf{V}_k$. To do so, we replace $\sigma$ in the AR(1) prior variance with the multiplicative gamma shrinkage prior. In practice, this model is fit by specifying a value of $K$ believed to be much larger than the true $K$ (which can be thought of as an upper bound on $K$), and the model automatically shrinks the unnecessary factors. The priors for this model can be mathematically formalized as follows:

$$
\begin{gathered}
    \mathbf{U}_{k} \sim \mathrm{N}(0,\tau[D_S-W_S]^{-1})\\
    V_{kt}\mid a,b, \phi_{kt}, \eta_{k}, V_{k(t-1)} \sim \mathrm{N}\left(a + b *V_{k(t-1)},\phi_{kt}^{-1} \eta_{k}^{-1}\right) 
\end{gathered}
$$
Here, improper flat priors (i.e. $\mathrm{Uniform}(-\infty,\infty)$) are placed on the coefficients, $a$ and $b$. $\phi_{kt}$ is a local shrinkage parameter for the elements in the $K$th row with a $\operatorname{Ga}(\nu / 2, \nu / 2)$ prior. $\eta_{k}=\prod_{l=1}^{k} \delta_{l}$ is a global shrinkage parameter for the $K$th row. The $\delta_{l}$ $(l=1, \ldots, \infty)$, are independent and $\delta_{1} \sim \mathrm{Ga}\left(a_{1}, 1\right)$ and $\delta_{l} \sim \mathrm{Ga}\left(a_{2}, 1\right)$.  The $\eta_{k}$s are stochastically increasing under the restriction $a_{2}>1$ as $K$ increases, which favors more shrinkage as the row index increases.

\subsection{Data pre-processing for rare outcomes}\label{ss:smoothing}

Due to the potentially extreme temporal instability in rare disease time series, we also consider applying temporal smoothing to the time series from each county as a pre-processing step, prior to their inclusion in the models. Pre-processing of noisy time series prior to formal analysis has long been applied in fields like neuroimaging \cite{ni_edu} to improve downstream model performance and ability to detect relevant signals. Here, we consider each of the models described above both with and without a temporal smoothing procedure implemented on data as a pre-processing step. When this pre-processing is applied, we fit a Poisson generalized linear model with a smoothed term on time to each of the time series individually, obtain ``smoothed'' predictions of the outcome at each time point from the models, and plug those into the MC approach in place of the raw data. That is, we fit

$$
\mathrm{log}(E[\mathbf{Y}_{it}]) = \mathrm{ns}(\mathrm{time}_t, 5) + \mathrm{log}(\theta_{it})
$$
on the observed data for each county where $\mathbf{Y}_i$ is the observed outcome vector for county $i$, $\mathrm{ns}(\mathrm{time}_t, 5)$ applies a natural spline with 5 degrees of freedom to the time variable, and $\mathrm{log}(\theta_{it})$ represents the population offset. We then obtain the ``smoothed'' predictions for county $i$ by applying the inverse of the link function to its estimated linear predictor.

\section{Simulations}
We conduct extensive simulation studies to evaluate our models' performance, both relative to one another and relative to the current gold standard approaches, in the context of outcomes mimicking rare and non-rare disease incidence. We compare the counterfactual estimation accuracy of our Bayesian MC methods, implemented in the \texttt{rstan} R package \cite{rstan}, and the each of the following existing MC methods: that of Athey et al. as implemented in the \texttt{gsynth} R package \cite{athey2021matrix,gsynth}, alternating least squares (ALS) from the \texttt{recommenderlab} R package \cite{takacs2012alternating, recommenderlab}, and soft impute, nuclear norm minimization, and singular value threshold (SVT) from the \texttt{filling} R package \cite{mazumder2010spectral, candes2012exact, cai2010singular, filling}. We also compare to the generalized synthetic controls method (GSC) from the \texttt{gsynth} package \cite{xu2017generalized,gsynth} and the augmented synthetic control method (ASC) from the \texttt{augsynth} R package \cite{Ben2021,augsynth}.

\subsection{Simulation Structure}
Panel data are simulated with $N=29$ counties and $T=15$ years, with $n_{\mathcal{W}}=6$ treated units initiating treatment at year $t=9$, and using a real spatial adjacency matrix for New Mexico's counties. We generate $K=3$ time varying latent factor vectors $\mathbf{V}_k$ and unit-specific factor loading vectors $\mathbf{U}_k$ with conditionally autoregressive forms as follows:
$\mathbf{U}_k \sim \mathrm{N}(\mathbf{0}_n,\tau^2\mathbf{Q}^{-1}\left(\mathbf{W}, \rho_{S}\right)) $ and $ \mathbf{V}_k \sim \mathrm{N}(\mathbf{0}_m,\tau^2\mathbf{Q}^{-1}\left(\mathbf{T}, \rho_{T}\right))$. By generating the data using only three latent factors ($K=3$), we ensure our data has the low-rank structure we would expect with rare cancer incidence data. Because cancer incidence data are known to exhibit spatial and temporal correlation \cite{francis2020spatial, gustafsson1999evidence, kreis2016space, mcnally2002space,gilman1995childhood,agost2016analysis}, we expect any time-varying unobserved confounders to as well. Thus, spatial and temporal correlation are induced in $\mathbf{U}_k$ and $\mathbf{V}_k$, respectively, through the precision matrices which are defined as in Leroux et al. (2000) \cite{leroux2000estimation}:

$$
\mathbf{Q}\left(\mathbf{W}, \rho_{S}\right)= \rho_{S}[\operatorname{diag}(\mathbf{W 1}_W)-\mathbf{W}]+\left(1-\rho_{S}\right) \mathbf{I}_N$$
$$\mathbf{Q}\left(\mathbf{T}, \rho_{T}\right)=\rho_{T}[\operatorname{diag}(\mathbf{T 1}_T)-\mathbf{T}]+\left(1-\rho_{T}\right) \mathbf{I}_T
$$ where $\mathbf{1}_W$ and $\mathbf{1}_T$ are $N \times 1$ and $T \times 1$ vectors of ones, respectively. Here, $\mathbf{W}$ and $\mathbf{T}$ are spatial and temporal adjacency matrices, respectively, and $\rho_S$ and $\rho_T$ are common spatial and temporal dependence parameters. In these simulations, we let $\rho_S=\rho_T=.99$. For the main simulations, the spatial and temporal variance parameters are set to $\tau^2=0.1$. To evaluate a setting where the spatio-temporal factor structure contributes less to the overall variability in the outcome, we also ran simulations with $\tau^2=0.08$.

The elements of the panel matrix, $\mathbf{Y(0)}$, are simulated from a Poisson distribution with mean $E\left[Y_{it}(0)\right]=\mathrm{exp}(\alpha + \gamma_i + \psi_t + \mathbf{U}_i^{T}\mathbf{V}_t +\mathrm{log}(\theta_{\mathrm{it}}))$. The $\theta_{\mathrm{it}}$ are the real population counts for children ages 0 to 9 in the New Mexico county $i$ for time point $t$, extracted from US Census Bureau data, and $\gamma_i, \psi_t \sim N(0,.000015)$. We modified $\alpha$ to obtain simulated data emulating non-rare ($\alpha = -5$) and rare ($\alpha = -7$) outcomes. In each of the rare and non-rare outcome settings, we generated 100 synthetic datasets from the above data generating mechanisms.

We applied Models 1-5 above to each synthetic dataset with 3 different specifications of $K$: one under-specified ($K=1$), one correct ($K=3$), one over-specified ($K=7$). For Model 6 above, we apply it as recommended by Bhattacharya and Dunson \cite{bhattacharya2011sparse}, specifying only a large value of $K$, $K=7$, in anticipation that it will adaptively shrink away unnecessary factors. 2,000 Hamiltonian Monte Carlo iterations with 1,000 burn-in iterations were collected for each model from \texttt{rstan} \cite{rstan}. For the competing existing methods, we used the rates themselves as the outcome, since these methods cannot straightforwardly accommodate an offset. We also considered the same three specifications of K for the Athey MC and ALS MC approaches, which require user-specification of $K$, whereas all other competing existing methods choose $K$ adaptively. Tuning parameters for Soft Impute MC and SVT MC were chosen via ten-fold cross-validation. Note that none of the models applied to the data is the ``true'' data generating model, so here we are evaluating the performance of all the models under some degree of model misspecification. For all of the approaches, we apply and evaluate them with and without the temporal smoothing pre-processing step described in Section~\ref{ss:smoothing}. We fit the smoothing models for each time series using a spline on time with 5 degrees of freedom. To measure performance, we compute the average absolute percent bias in the estimates of the estimated counterfactuals for each approach.

\subsection{Results}

\begin{table}[H]
\begin{center}

\caption{Simulation results: Average absolute percent bias of the counterfactual estimates from each method applied with different specifications of the number of factors ($K$) where applicable (rows), in simulated data emulating rare and non-rare outcomes, with and without a temporal smoothing pre-processing step (columns).}
\label{tab:absPercBiasY0}
\begin{tabular}{ 
p{4.6cm}cC{2.5cm}C{2.2cm}C{2.5cm}C{2cm} } 
\toprule
\multirow{2}{*}{Model} & \multirow{2}{*}{ K } & Non-Rare \&  Non-Smoothed  & Non-Rare \& Smoothed & Rare \& Non-Smoothed & Rare \& Smoothed\\
\toprule
\multicolumn{6}{l}{\textbf{Bayesian MC Models}} \\ \bottomrule
\multirow{3}{6em}{\hspace{.4cm}1. Vanilla} & 1 & 11.56 & 11.11 & 29.65 & 29.87 \\
 & 3 & \textbf{10.50} & 11.33 & 32.77 & 30.90 \\

 & 7 & 11.12 & 11.64 & 33.89 & 34.18\\
\hline
\multirow{3}{6em}{\hspace{.4cm}2. Space} & 1 & \textbf{10.29} & \textbf{10.50} & 29.60 & \textbf{28.51}\\

 & 3 & \textbf{10.69} & \textbf{10.45} & \textbf{29.38} & \textbf{28.33}\\

 & 7 & 10.79 & \textbf{10.53} & \textbf{28.62} & 28.96\\
\hline
\multirow{3}{*}{\hspace{.4cm}3. Space-Time ICAR} & 1 & 10.90 & 10.69 & 30.42 & \textbf{28.04}\\

 & 3 & 10.83 & 10.66 & 29.45 & 29.89\\

 & 7 & 10.76 & 10.56 & \textbf{29.18} & 29.33\\
\hline
\multirow{3}{*}{\hspace{.4cm}4. Space-Time AR} & 1 & 10.87 & \textbf{10.46} & 30.46 & 29.94\\

 & 3 & \textbf{10.32} & 10.56 & \textbf{28.94} & \textbf{28.87}\\

 & 7 & 10.95 & \textbf{10.29} & \textbf{28.99} & \textbf{28.00}\\
\hline
\multirow{3}{*}{\hspace{.4cm}5. Space-Time Lasso} & 1 & 13.35 & 12.32 & 31.02 & 30.01\\

 & 3 & 12.59 & 13.19 & 37.86 & 37.26\\

 & 7 & 16.30 & 16.47 & 57.03 & 46.27\\
\hline
\hspace{.4cm}6. Space-Time Shrinkage & 7 & \textbf{10.66} & 10.65 & 29.94 & 29.38\\
\toprule
\multicolumn{6}{l}{\textbf{Existing Methods}} \\ \bottomrule
\multirow{3}{6em}{\hspace{.4cm}Athey MC} & 1 & 13.78 & 13.89 & 32.34 & 34.98\\

 & 3 & 14.14 & 14.43 & 32.22 & 34.70\\

 & 7 & 14.75 & 14.49 & 31.73 & 33.89\\
\hline
\multirow{1}{6em}{\hspace{.4cm}GSC} & NA & 14.11 & 14.13 & 31.92 & 34.70\\
\hline
\multirow{1}{6em}{\hspace{.4cm}ASC} & NA & 14.04 & 13.13 & 36.14 & 33.28\\
\hline
\multirow{3}{6em}{\hspace{.4cm}ALS MC} & 1 & 12.14 & 12.48 & 34.25 & 33.01\\

 & 3 & 14.80 & 13.78 & 37.97 & 35.06\\

 & 7 & 21.90 & 19.92 & 52.46 & 42.89\\
\hline
\multirow{1}{*}{\hspace{.4cm}Soft Impute MC} & NA & 99.33 & 99.33 & 96.14 & 96.21\\
\hline
\multirow{1}{*}{\hspace{.4cm}Nuclear Norm MC} & NA & 12.99 & 12.89 & 34.58 & 37.06\\
\hline
\multirow{1}{6em}{\hspace{.4cm}SVT MC} & NA & 36.35 & 36.24 & 41.68 & 40.26\\
\hline
\end{tabular}
\end{center}
\end{table}

Table~\ref{tab:absPercBiasY0} shows the average bias results for the primary simulations.  For more detail on the distribution of the biases across simulations, see the 25$^{\mathrm{th}}$ and 75$^{\mathrm{th}}$ percentiles in Table \ref{tab:iqr}. The five bold numbers in each column represent the models with the lowest percent biases in the given simulation scenario. We first note that, in almost all cases, the magnitude of bias is 2-3 times larger for any given method when applied to rare outcomes compared to non-rare ones. As expected, this confirms that all of these approaches, which rely on stable trends to properly adjust for confounding bias, struggle more in the presence of rare disease outcomes like CYA cancer incidence.

For all of our Bayesian MC models, we generally see lower bias in the counterfactual estimates when comparing to the existing methods. The existing methods that performed best were ALS and Nuclear Norm MC with non-rare outcomes and Athey MC with rare outcomes. 

Overall, across simulation scenarios and choices of $K$, we see the best and most consistent performance with the Space MC Model and the Space-Time AR MC Model. In each simulation scenario, the Space and Space-Time AR models (under various specifications of $K$) account for at least three out of the five best-performing models. The Space-Time ICAR MC model also ranks among the top performers in the rare settings.

In general, the a priori temporal smoothing step does not seem to improve performance. The choice of $K$ also also has relatively little influence on performance for most models, indicating robustness of the models to misspecification of $K$. The space-time shrinkage model, while not consistently among the top five performers, is generally quite competitive with the best methods and may be advantageous in practice because it only requires specification of an upper bound on $K$.

The results from the simulations using lower value of $\tau^2$ in the data generating process are shown in Table \ref{tab:absPercBiasY0_alttau}. Overall the results were similar, but SVT MC became more competitive with Athey MC and the Bayesian methods. This indicates that, not surprisingly, existing methods may be adequate in settings where the contribution of the spatio-temporal components to the overall variance in the outcome is low.

\section{Application}\label{s:application}

We applied each of the Bayesian MC models and the Athey et al. MC implementation in the \texttt{gsynth} package to the data for each disease type separately with and without prior temporal smoothing, while controlling for percent Hispanic. That is, $\mathrm{X}_{it}$ in equation (2) is the percent of the population that is Hispanic for county $i$ and time $t$. In a preliminary principal components analysis (Figures \ref{fig:ScreeLymphoma} and \ref{fig:ScreeLeukemia}), we found for both data sets most of the variation is explained with two principal components. This suggests that a small number of latent factors effectively explains the variance within the datasets, indicating that the assumption of a low-rank factorization is reasonable in this context. Because of this, $K=2$ was chosen for the Vanilla, Space, Space-Time ICAR, Space-Time AR, Space-Time Lasso, and Athey MC Models. $K=3$ was chosen for the Space-Time Shrinkage Model. For each of our models, 2,000 Hamiltonian Monte Carlo iterations were run with 1,000 burn-in iterations in \texttt{rstan}. We report estimated ATTs for each disease type averaged over all treated counties and times, as well as averaged separately for each of the two treated states (California and Connecticut) and averaged separately for each time point.

Table \ref{tab:rhat} displays the average R-hat value, a Bayesian model convergence diagnostic \cite{vehtari2021rank}, for each model's counterfactual predictions. An R-hat of less than 1.05 for a given parameter suggests that its chain of posterior samples has converged. Each of our models resulted in average R-hat values for the counterfactual predictions of less than 1.05, indicating adequate convergence of our parameters of interest. 

The overall and state-specific estimated ATTs and 95\% credible intervals (CIs) for each method, in the absence of temporal smoothing, are given in Table \ref{tab:ATTLymphomaLeukemia} for childhood lymphoma and leukemia. Table \ref{tab:ATTLymphomaLeukemiaSmooth} displays the results with the temporal smoothing pre-processing step. These ATTs were calculated by computing the time specific ATTs, $\mathrm{ATT}_t^{(m)}$, and then averaging across times. The ATTs were uniformly negative for lymphoma, with negative values indicating reductions in lymphoma incidence rates relative to the rates expected in the counterfactual scenario where no reformulated gasoline program took place. The overall ATT estimate from the Space-Time AR model, our preferred model in the simulations, was -1.00 (-1.40, -0.57). This estimate can be interpreted as follows: the rate of childhood and young adult lymphoma decreased in California and Connecticut counties by an average of 1 case per 100,000 due to the reformulated gasoline program. None of the 95\% CIs for the overall ATTs contained zero (the null value), providing strong evidence that the reformulated gasoline program led to a decrease in childhood lymphoma incidence. We also observed more strongly negative ATTs in Connecticut counties compared to California counties.

The results for leukemia were inconclusive, as ATT estimates were largely positive but with wide CIs generally including zero (Table \ref{tab:ATTLymphomaLeukemia}). For instance, the overall ATT estimate from the Space-Time AR model was 0.05 (-0.61, 0.64). CIs containing zero were also generally observed for California and Connecticut separately. Although we could in theory estimate county-specific effects from our models as is done in \cite{nethery2021integrated}, we argue that we should not do so in the rare outcomes setting as the county-specific estimates are often too unstable.

\begin{table}[H]
\caption{ Overall and state-specific average treatment effect on the treated (ATT) estimates and 95\% confidence/credible intervals (CIs) quantifying the effect of the reformulated gasoline program on childhood lymphoma and leukemia incidence.}
\label{tab:ATTLymphomaLeukemia}
\begin{tabular}{l c c c c c c}

\toprule
\multirow{2}{*}{ Model }  & \multicolumn{2}{c}{ California } & \multicolumn{2}{c}{ Connecticut } & \multicolumn{2}{c}{ Overall }\\
 & ATT & 95\% CI & ATT & 95\% CI & ATT & 95\% CI\\
\toprule 
\multicolumn{7}{l}{\textbf{Lymphoma}}\\
\toprule 
Athey MC\tablefootnote{The implementation of Athey's MC in \texttt{gsynth} only includes confidence intervals for the overall ATT. State specific ATTs were calculated using the resulting estimated counterfactual values.} & -0.46 & NA & -1.60 & NA & -1.16 & (-2.22 , -0.10)\\

Vanilla & -0.94 & (-2.11 , -0.04) & -1.81 & (-2.95 , -0.80) & -1.46 & (-2.42 , -0.70)\\

Space & -1.11 & (-2.83 , 0.04) & -1.77 & (-3.04 , -0.69) & -1.54 & (-2.63 , -0.68)\\

Space-Time 
 ICAR & -0.83 & (-1.97 , 0.12) & -1.71 & (-2.94 , -0.70) & -1.38 & (-2.29 , -0.62)\\

Space-Time 
 AR & -0.80 & (-1.29 , -0.27) & -1.13 & (-1.73 , -0.55) & -1.00 & (-1.40 , -0.57)\\

Space-Time 
 Lasso & -0.96 & (-1.88 , -0.15) & -1.73 & (-2.62 , -0.79) & -1.44 & (-2.14 , -0.71)\\

Space-Time 
 Shrinkage & -0.64 & (-1.35 , 0.09) & -1.46 & (-2.28 , -0.62) & -1.16 & (-1.70 , -0.57)\\

\toprule
\multicolumn{7}{l}{\textbf{Leukemia}}\\
\toprule
Athey MC & 0.36 & NA & 4.14 & NA & 0.92 & (-0.74 , 2.57)\\

Vanilla & 0.45 & (-0.92 , 1.55) & 1.26 & (-0.59 , 2.73) & 0.58 & (-0.71 , 1.54)\\

Space & 0.84 & (-0.39 , 1.61) & 1.13 & (-0.32 , 2.27) & 0.86 & (-0.23 , 1.57)\\

Space-Time 
 ICAR & 0.75 & (-0.39 , 1.71) & 1.31 & (0.08 , 2.36) & 0.84 & (-0.17 , 1.69)\\

Space-Time 
 AR & -0.13 & (-0.88 , 0.49) & 1.13 & (0.18 , 2.06) & 0.05 & (-0.61 , 0.64)\\

Space-Time 
 Lasso & 0.60 & (-0.50 , 1.49) & 1.32 & (-0.50 , 2.62) & 0.68 & (-0.35 , 1.54)\\

Space-Time 
 Shrinkage & 0.26 & (-0.99 , 1.47) & 1.27 & (-0.14 , 2.64) & 0.38 & (-0.78 , 1.57)\\
\toprule
\end{tabular}

\end{table}

\begin{figure}[h!]
\centering
\subfloat[Lymphoma]
{\includegraphics[width=1\linewidth]{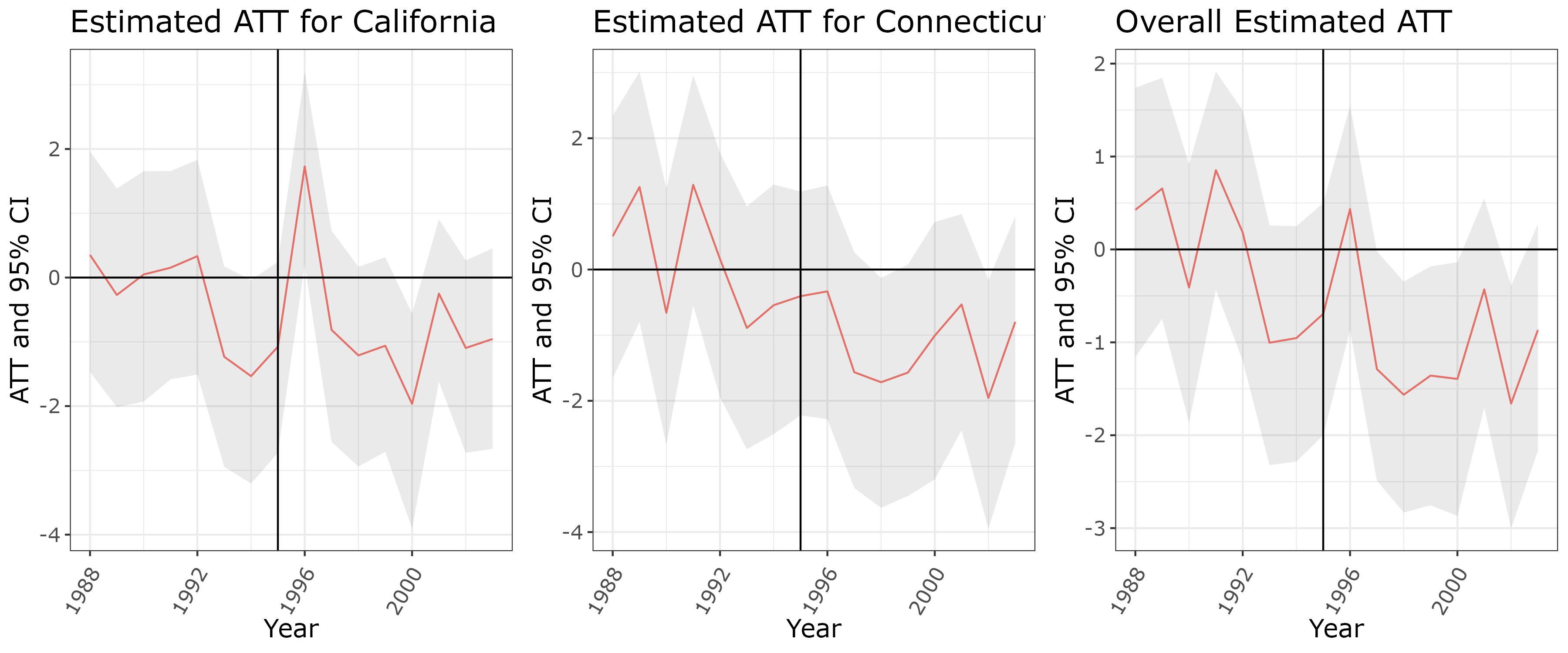}\label{fig:ATT_time_lymph}}

\subfloat[Leukemia]
{\includegraphics[width=1\linewidth]{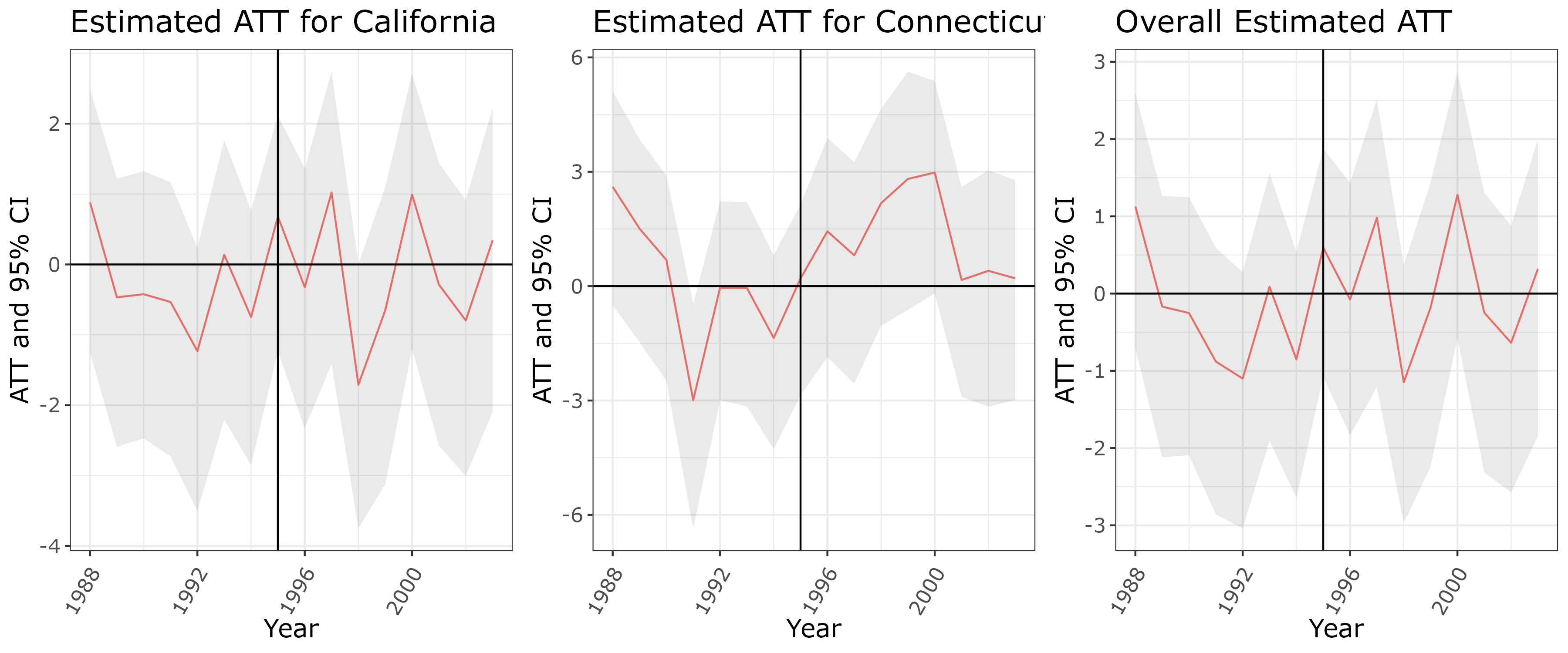} \label{fig:ATT_time_leuk}}

\caption[Estimated ATT Over Time]{Plots showing the estimated ATT over-time for incidence of lymphoma (a) and leukemia (b). Space-Time AR Models were chosen based on evaluation of all models pre-treatment fits. \label{fig:ATT_time}}
\end{figure}

Figure \ref{fig:ATT_time} shows the ATT estimates and CIs at each time point, averaged over all treated counties, for lymphoma and leukemia from the Space-time AR models. Here, we show the pre-treatment ATT estimates as a diagnostic to investigate how well the model's latent factors capture key time-varying features that influence the outcome in the absence of treatment. Pre-treatment ATT estimates centered around zero suggest good model fit. Systematic temporal trends in the pre-treatment ATT can indicate (a) that some important time-varying features are not being captured by the latent factors or (b) the presence of anticipation effects of treatment in treated areas. For lymphoma, (Figure \ref{fig:ATT_time_lymph}), pre-treatment ATT estimates are largely fluctuating around zero, with possible evidence of a decreasing trend beginning around 1993, and ATT estimates continue decreasing post-treatment. For leukemia (Figure \ref{fig:ATT_time_leuk}), there are no apparent systematic pre-treatment temporal trends in the ATT estimates. 

The reformulated gasoline program was first passed and announced in 1990, although it was not imposed in covered areas until 1995. However, prior research suggests that there was an anticipatory effect of the program on gasoline content and TRAP in covered areas, which may explain the declining lymphoma ATT estimates prior to treatment. Take for example benzene, which is a key traffic-related pollution that has been implicated in hematologic cancer development and one that was targeted for reduction in reformulated gasoline \cite{rinsky1981leukemia,rinsky1987benzene}. In California, ambient benzene concentrations at monitoring sites was decreasing steadily during the period 1990-1994, prior to the formal implementation of the reformulated gasoline program, with a total decrease of about 40\% during this pre-treatment period \cite{propper2015ambient}. Although there are no benzene monitors in Connecticut, ambient benzene concentrations in nearby New York City began declining in 1990, although there was a clear, sharp decline between 1995 and 1996 \cite{aleksic2005concentrations}. It is possible that since ambient benzene (and possibly other traffic-related pollutants) began declining prior to the implementation of the reformulated gasoline program, any resulting health impacts might have begun to materialize prior to 1995. 

Because of this possible anticipatory effect of the reformulated gasoline program, we conduct sensitivity analyses considering 1990 as the treated year, and expanding our study period to 1983-2003 to allow for adequate pre-treatment data for these analyses (Section \ref{sec:1990trt}). The pre-treatment ATT estimates from these models are centered around zero and absent systematic temporal trends for both leukemia and lymphoma (Figure \ref{fig:ATT_time_Supp}), indicating strong model fit for both diseases. However, the results are very similar to those in our main analyses, with identical conclusions about the effects of the reformulated gasoline program. Tables \ref{tab:SuppATTLymphomaLeukemia_SensAnalysis}  - \ref{tab:SuppATTLymphomaLeukemia_SensAnalysis_smooth} show that the program led to decreases in lymphoma incidence, but results remain inconclusive for leukemia incidence. 

We also conducted sensitivity analyses with increased $K$ ($K=3$ for Vanilla, Space, Space-Time ICAR, Space-Time AR, Space-Time Lasso, and Athey MC models and $K=4$ for the Space-Time Shrinkage model). Results are shown in Tables \ref{tab:ATTLymphomaLeukemiaSmooth} and \ref{tab:ATT_SensAnalysis_K} and are similar to the results of our main analyses, indicating little sensitivity of our findings to choice to $K$. 

Recall that in MC for causal inference, the ATT is generally computed by plugging in the observed $Y_{it}=Y_{it}(1)$ for the treated units post-treatment \cite{athey2021matrix}. To assess the possible impact of noise in these observed outcomes on our ATT estimates, we conducted sensitivity analyses in which we spatio-temporally smoothed these outcomes and plugged in the smoothed values in place of the observed ones in the computation of the ATT. More details and the table of the results are provided in Section ~\ref{s:modeled_Y1}. We again found that our results were robust. 

\section{Discussion}

In this paper, we propose and evaluate causal inference approaches for studying the effects of a quasi-experiment on rare disease outcomes. Working within a Bayesian MC framework, we consider a number of prior distribution configurations to allow for spatio-temporal smoothing, regularization, and adaptive hyperparameter selection to reduce over-fitting and improve estimation in rare outcome contexts. The Bayesian approach also lends itself to more straightforward uncertainty quantification than its frequentist or machine learning counterparts \cite{tanaka2021bayesian}. Through our simulation study, we show that our proposed models perform better than numerous existing methods for quasi-experimental analysis in the context of rare disease outcomes. We find that both independent prior distributions and certain spatio-temporal priors on the MC latent factors and loadings perform well, and that over-specifying the unknown number of latent factors presents a greater threat to predictive accuracy than under-specifying when working with disease outcomes.  

Building on the sparse infinite factor model framework of Bhattacharya and Dunson \cite{bhattacharya2011sparse}, we also introduce and implement prior distributions on the latent factors and factor loadings that can simultaneously perform spatio-temporal smoothing and adaptively select the number of latent factors. This eliminates the need for user-specification of $K$, which may reduce model sensitivity to arbitrary tuning parameter selection in real data applications. To our knowledge, such prior distributions have not yet been proposed in the MC literature. Our simulation studies showed that this approach consistently performed well, with results comparable to the best-performing models under user-specification of $K$.

In our real data analysis, we found evidence that the EPA's reformulated gasoline program may have led to reductions in childhood and young adult lymphoma, but had no effect on childhood leukemia. These findings are consistent with the recent findings of Nethery et al. \cite{nethery2023mobile}, who studied the effects of a separate, smaller-scale gasoline reformulation program implemented by the US EPA in Alaska in 2011. This program, known as the Mobile Source Air Toxics Rule, was specifically focused on limiting the amount of benzene in gasoline. Applying difference-in-differences methods, Nethery et al. also observed a reduction in childhood and young adult lymphoma incidence following the implementation of this program (but no impact on childhood leukemia incidence). While prior literature based on observational studies has primarily reported associations between TRAP exposure and childhood leukemia \cite{raaschou2006air,boothe2014residential,sun2014no,filippini2015review}, these recent findings of associations between TRAP and childhood lymphoma using more robust quasi-experimental study designs suggest that greater attention to this potential link is warranted. 

A limitation of our work is that we do not have information about cross-county migration, which could induce exposure measurement error. For example, a person could develop leukemia or lymphoma while living in an unregulated county and then move to a regulated county where diagnosis occurs (or vice versa). Additionally, violations of the standard SUTVA assumptions for causal inference are possible in our real data application. For instance, SUTVA requires that there is only one ``version'' of treatment. Here, the intervention likely led to different degrees of change in exposure (likely larger decreases in TRAP exposure in urban areas vs. rural). However, because MC approaches do not model the outcomes under treatment (instead simply making use of the observed outcomes under treatment to estimate causal effects), we anticipate that any such issues would not affect our modeling and would only compel a more nuanced interpretation of the causal quantities being estimated. SUTVA also requires the ``no interference'' condition, i.e., that the treatment of one unit should not affect the potential outcomes of another unit. Violations of no interference could occur, for instance, if residents of unregulated areas commute regularly to nearby regulated areas and thereby reap the health benefits of the improved air quality in regulated areas. However, the structure of the reformulated gasoline program implementation and our SEER data are likely to mitigate this issue, as entire states are considered treated or untreated, and our data do not include any bordering states. Future work may expand on spatio-temporal matrix completion to include network effects as done in \cite{agarwal2022network}. This expansion could lead to a more comprehensive understanding of the underlying dynamics, thus enhancing the predictive accuracy and applicability of such models in various domains.

Given recent decreased focus on the development of federal environmental regulations in the US, individual cities and states have begun to make their own laws to decrease environmental exposures suspected to be harmful \cite{safestates}. These localized environmental health regulations are ideal QEs and provide the opportunity to study the effects of many different contaminants on disease outcomes. The methods introduced here will enable robust analysis of the health impacts of these QEs, including impacts on rare disease outcomes.

\section*{Acknowledgements}

Support for this research was provided by NIH grant K01ES032458. The collection of cancer incidence data used in this study was supported by the California Department of Public Health pursuant to California Health and Safety Code Section 103885; Centers for Disease Control and Prevention's (CDC) National Program of Cancer Registries, under cooperative agreement 1NU58DP007156; the National Cancer Institute's Surveillance, Epidemiology and End Results Program under contract HHSN261201800032I awarded to the University of California, San Francisco, contract HHSN261201800015I awarded to the University of Southern California, and contract HHSN261201800009I awarded to the Public Health Institute. The ideas and opinions expressed herein are those of the author(s) and do not necessarily reflect the opinions of the State of California, Department of Public Health, the National Cancer Institute, and the Centers for Disease Control and Prevention or their Contractors and Subcontractors.

\section*{Data and code availability}

The cancer incidence data used here cannot be redistributed under the conditions of the Data Use Agreements. However, interested parties can request the data through the standard processes described on the SEER and California Cancer Registry websites. Code to reproduce all simulations and real data analyses is available at \url{https://github.com/sofiavega98/trap_cancer_SpaceTimeMC/}.

\newpage
\bibliographystyle{unsrt}
\bibliography{bib}

\newpage
\pagenumbering{arabic}
\renewcommand*{\thepage}{S\arabic{page}}
\beginsupplement
\section{Supplementary Materials}
\setcounter{figure}{0}
\setcounter{table}{0}

\subsection{Causal Identifying Assumptions}\label{sec:assumptions}
In our paper, we define potential outcomes using Stable Unit Treatment Value Assumption (SUTVA) from causal inference \cite{rubin1980randomization}. This assumption posits two key principles: firstly, there exists only one version of treatment, and secondly, the outcome for any given unit remains unaffected by the treatment statuses of other units. 

When defining a unit's potential outcomes as a function at time $t$ solely based on its treatment status at that time, as expounded in the primary text, we implicitly make two standard assumptions. Firstly, we presume that future treatments cannot influence past outcomes. Secondly, we assume that a unit's entire historical treatment status vector is a deterministic function of the unit's initial time of treatment adoption. That is, treated units initiate treatment at some time $T_0$ and remain treated through time $T$. Consequently, a unit's potential outcomes can be expressed solely as functions of its time of initial treatment adoption, rather than functions of its complete historical treatment status vector.

In addition to the above framework, we adhere to the following four assumptions necessary for identifying causal effects using the matrix completion approach presented:\\

\textit{Assumption 1: Causal Consistency} - This posits that the observed outcome equals the potential outcome under the observed treatment level.\\

\textit{Assumption 2: Latent Ignorability} - This states that, given observed covariates and a set of latent variables, the treatment assignment is independent of potential outcomes under control.\\

\textit{Assumption 3: Approximation of Unobservables} - This assumes that unit-specific vectors of unobserved features can be approximated through a low-rank latent factor model.\\

\textit{Assumption 4: Conditional Exchangeability} - Conditional on observed covariates and latent variables, elements of $\mathbf{Y(0)}$ are exchangeable, facilitating the definition of the posterior predictive distribution used to estimate missing values in Y(0).

See \cite{tanaka2021bayesian,nethery2021integrated,pang2021bayesian} for further details on the above assumptions.

\subsection{Computation}\label{sec:computation}

The seminal paper on spatial modeling with \texttt{rstan} noted that numerical issues can occur when drawing samples from the posterior predictive distribution in a given iteration \cite{morris2019bayesian}, which can cause the entire model run to fail. They recommended implementing a numerical constraint in the \texttt{rstan} function that will prevent it from throwing an error if this occurs, which we also employ here. Specifically, the constraint says that, if in any iteration $m$, $\text{log}(E[Y_{it}^{(m)}(0)]) > 20$ for any $(i,t)$, then we should not try to draw a posterior predictive sample and move on to the next iteration (in practice, the issue is flagged by setting the posterior predictive samples for that particular iteration equal to -1 in the output). Although this approach allows the model to continue running and produce results, it is important to proceed with caution if this occurs, as it does indicate model instability and the results might not be reliable. In our simulation section, only the Space-Time Lasso model applied to the rare data experienced these numerical issues when run on some of the simulated datasets. We included the results of those simulations when summarizing the biases in Table~\ref{tab:absPercBiasY0}, which likely partly explains the poor performance of the Space-Time Lasso Model.

Additionally, to speed computation, we utilize a soft-sum to zero constraint on the factors of $\mathbf{U}$ and $\mathbf{V}$ for all models besides the Vanilla Model. That is,
we apply a $\text{N}(0,0.01)$ prior to $\text{mean}(\mathbf{U}_k)$ and $\text{mean}(\mathbf{V}_k)$ as suggested by \cite{morris2019bayesian}.

\subsection{Real Data Sensitivity Analysis}\label{sec:1990trt}

Since ambient benzene began declining in 1990, we considered leukemia and lymphoma incidence data from 1983 to 2003 with 1990 as the treated year. Due to data access limitations, we used SEER incidence data for the both lymphoma and leukemia.

\begin{figure}[H]
  \centering
  \subfloat[Lymphoma]{\includegraphics[width=0.5\textwidth]{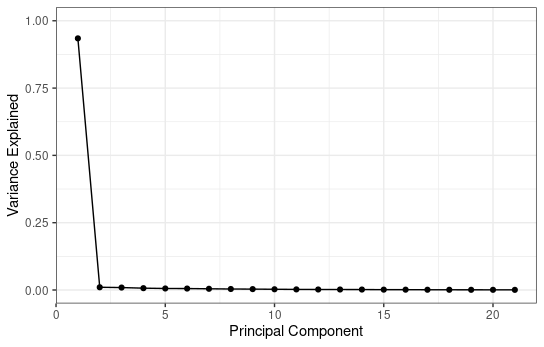}\label{fig:1990trt_ScreeLymphoma}}
  \hfill
  \subfloat[Leukemia]{\includegraphics[width=0.5\textwidth]{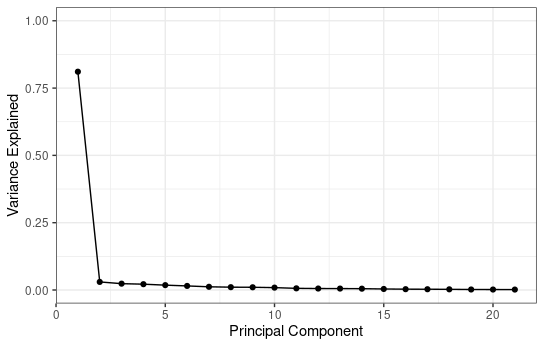}\label{fig:1990trt_ScreeLeukemia}}
  \caption{Scree plots showing the percent of variance explained by each principal component for each panel data matrix: lymphoma (a) and leukemia (b).}
  \label{fig:1990trt_ScreeLymphomaLeukemia}
\end{figure}

Figure \ref{fig:1990trt_ScreeLymphomaLeukemia} show preliminary principal component analyses for both data sets. With these results, $K=2$ chosen for the Vanilla, Space,
Space-Time ICAR, Space-Time AR, Space-Time Lasso, and Athey MC Models and $K=3$ was chosen for the Space-Time Shrinkage Model.

\begin{figure}[H]
\centering
\subfloat[Lymphoma]
{\includegraphics[width=1\linewidth]{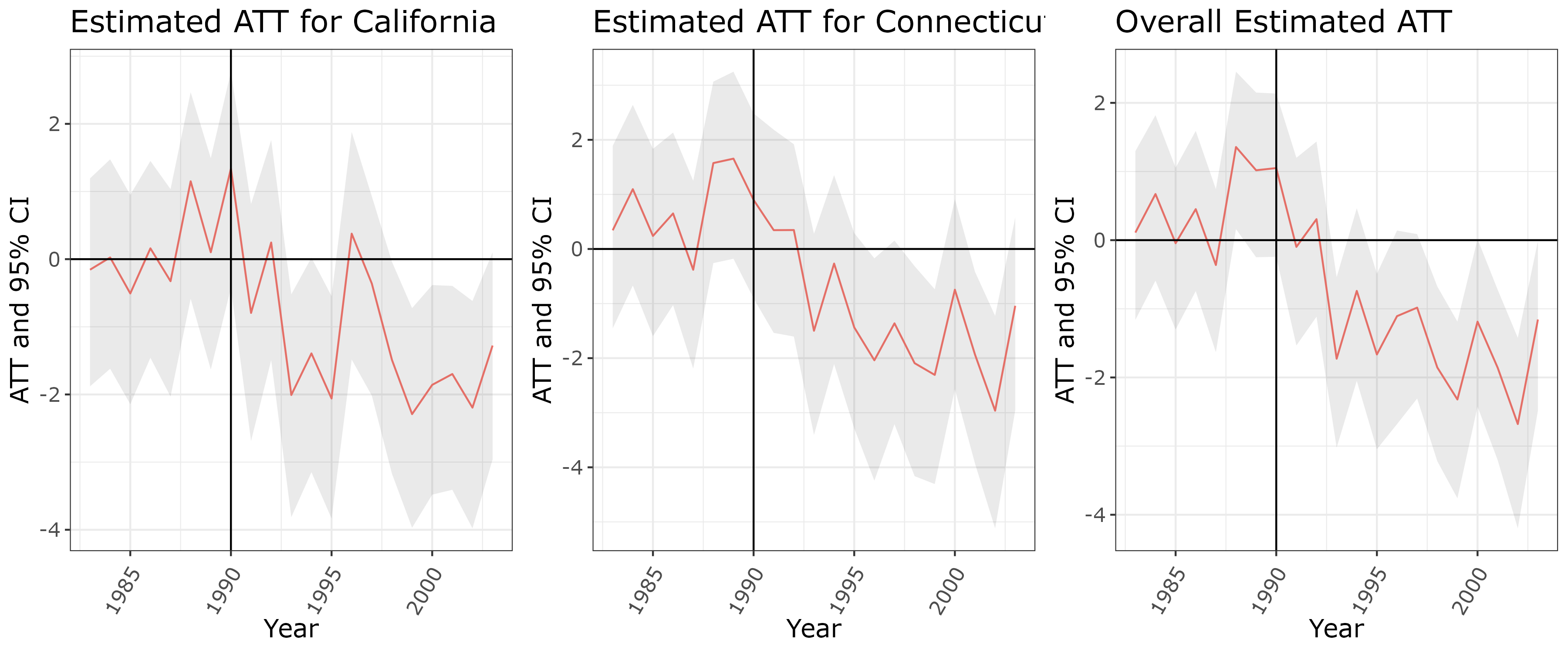}\label{fig:ATT_time_lymph_1990_Supp}}

\subfloat[Leukemia]
{\includegraphics[width=1\linewidth]{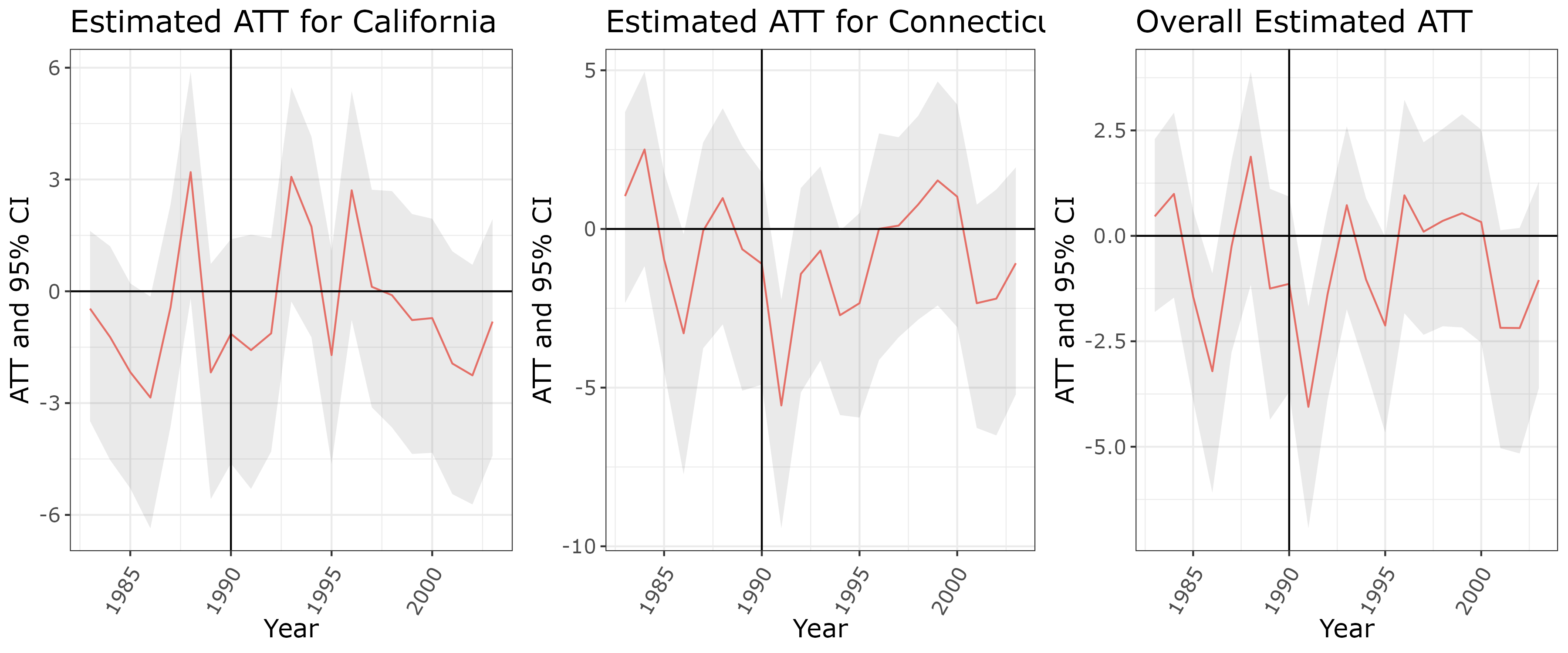} \label{fig:ATT_time_leuk_Supp}}

\caption[Estimated ATT Over Time]{Plots showing the estimated ATT over-time for incidence of lymphoma (a) and leukemia (b) in the absence of temporal smoothing. The Space-Time AR Model was chosen based on the results of the simulation study. \label{fig:ATT_time_Supp}}
\end{figure}

We similarly run 2,000 Hamiltonian Monte Carlo iterations with 1,000 burn-in iterations in \texttt{rstan} for each of our models. Table \ref{tab:rhat_SensAnalysis_trt1990} shows R-hat values for our models indicating proper convergence. Results are shown in Tables \ref{tab:SuppATTLymphomaLeukemia_SensAnalysis} and \ref{tab:SuppATTLymphomaLeukemia_SensAnalysis_smooth}. Figure \ref{fig:ATT_time_Supp} shows the estimated ATT over time. Overall results agree with those shown in the Application Section (Section \ref{s:application}), but \ref{fig:ATT_time_lymph_1990_Supp} shows improved pre-treatment trends for incidence of CYA lymphoma.

\begin{table}[H]
\caption{Average R-hat for each model's counterfactual predictions applied with different specifications of the number of factors, $K$ (rows), in CYA lymphoma and leukemia incidence data, with and without a temporal smoothing pre-processing step (columns).\label{tab:rhat_SensAnalysis_trt1990}}
\begin{center}

\begin{tabular}{ p{4cm}cC{2.5cm}C{2.5cm}C{2.5cm}C{2.5cm} } 
\toprule
\multirow{2}{*}{Model} & \multirow{2}{*}{ K } & Lymphoma \&  Non-Smoothed  & Lymphoma \& Smoothed & Leukemia \& Non-Smoothed & Leukemia \& Smoothed\\
\toprule
Vanilla & 2 & 1.00 & 1.00 & 1.00 & 1.00\\
Space & 2 & 1.00 & 1.00 & 1.01 & 1.03\\
Space-Time ICAR & 2 & 1.01 & 1.01 & 1.02 & 1.04\\
Space-Time AR & 2 & 1.00 & 1.00 & 1.00 & 1.00\\
Space-Time Lasso & 2 & 1.03 & 1.06 & 1.03 & 1.06\\
Space-Time Shrinkage & 3 & 1.02 & 1.01 & 1.02 & 1.01\\
\toprule
\end{tabular}

\end{center}

\end{table}

\begin{table}[H]
\caption{Overall and state-specific estimated ATT results for each model applied to CYA lymphoma and leukemia incidence data without a temporal smoothing pre-processing step.}
\label{tab:SuppATTLymphomaLeukemia_SensAnalysis}
\begin{tabular}{l c c c c c c}
\toprule
\multirow{2}{*}{ Model }  & \multicolumn{2}{c}{ California } & \multicolumn{2}{c}{ Connecticut } & \multicolumn{2}{c}{ Overall }\\

& ATT & 95\% CI & ATT & 95\% CI & ATT & 95\% CI\\
\toprule 
\multicolumn{7}{l}{\textbf{Lymphoma}}\\
\toprule 
Athey MC & -0.23 & NA & -1.08 & NA & -0.76 & (-1.59 , 0.08)\\

Vanilla & -0.83 & (-1.82 , -0.04) & -1.49 & (-2.65 , -0.52) & -1.24 & (-2.12 , -0.53)\\

Space & -0.70 & (-2.01 , 0.33) & -1.46 & (-3.28 , -0.32) & -1.17 & (-2.52 , -0.32)\\

Space-Time 
 ICAR & -0.60 & (-1.57 , 0.42) & -1.20 & (-2.43 , -0.08) & -0.94 & (-1.89 , -0.03)\\

Space-Time 
 AR & -1.13 & (-1.57 , -0.74) & -1.18 & (-1.68 , -0.71) & -1.17 & (-1.53 , -0.81)\\

Space-Time 
 Lasso & -0.82 & (-1.68 , -0.26) & -1.36 & (-2.18 , -0.66) & -1.16 & (-1.74 , -0.65)\\

Space-Time 
 Shrinkage & -0.65 & (-1.15 , -0.17) & -1.58 & (-2.29 , -0.79) & -1.22 & (-1.74 , -0.73)\\

\toprule 
\multicolumn{7}{l}{\textbf{Leukemia}}\\
\toprule

Athey MC & 3.52 & NA & 2.29 & NA & 2.76 & (1.28 , 4.25)\\

Vanilla & 0.80 & (-1.40 , 2.35) & -0.98 & (-3.09 , 0.75) & -0.33 & (-2.03 , 1.05)\\

Space & -0.98 & (-80.21 , 1.45) & -1.36 & (-5.54 , 0.49) & -1.32 & (-32.99 , 0.32)\\

Space-Time 
 ICAR & 0.22 & (-2.97 , 1.88) & -1.17 & (-4.05 , 0.33) & -0.68 & (-3.18 , 0.69)\\

Space-Time 
 AR & -0.41 & (-1.35 , 0.41) & -1.24 & (-2.21 , -0.34) & -0.92 & (-1.62 , -0.28)\\

Space-Time 
 Lasso & -7.74 & (-137.10 , -0.08) & -1.07 & (-4.86 , 0.94) & -3.67 & (-55.82 , 0.04)\\

Space-Time 
 Shrinkage & 0.91 & (0.15 , 1.67) & -0.12 & (-1.10 , 0.80) & 0.26 & (-0.41 , 0.91)\\

\toprule
\end{tabular}

\end{table}

\begin{table}[H]
\caption{Overall and state-specific estimated ATT results for each model applied to CYA lymphoma and leukemia incidence data with a temporal smoothing pre-processing step.}
\label{tab:SuppATTLymphomaLeukemia_SensAnalysis_smooth}
\begin{tabular}{l c c c c c c}
\toprule
\multirow{2}{*}{ Model }  & \multicolumn{2}{c}{ California } & \multicolumn{2}{c}{ Connecticut } & \multicolumn{2}{c}{ Overall }\\
& ATT & 95\% CI & ATT & 95\% CI & ATT & 95\% CI\\
\toprule 
\multicolumn{7}{l}{\textbf{Lymphoma}}\\
\toprule
Athey MC & 0.01 & NA & -0.70 & NA & -0.43 & (-1.15 , 0.43)\\

Vanilla & -0.68 & (-2.18 , 0.48) & -1.36 & (-2.96 , -0.22) & -1.14 & (-2.16 , -0.23)\\

Space & -0.81 & (-3.30 , 0.58) & -1.45 & (-3.64 , -0.19) & -1.23 & (-3.03 , -0.19)\\

Space-Time 
 ICAR & -1.03 & (-3.27 , 0.19) & -1.30 & (-2.85 , -0.08) & -1.25 & (-2.64 , -0.14)\\

Space-Time 
 AR & -0.40 & (-0.79 , -0.02) & -1.36 & (-1.86 , -0.89) & -0.99 & (-1.33 , -0.68)\\

Space-Time 
 Lasso & -0.96 & (-1.96 , -0.34) & -1.01 & (-1.83 , -0.30) & -1.00 & (-1.50 , -0.58)\\

Space-Time 
 Shrinkage & -0.59 & (-1.46 , 0.06) & -1.07 & (-1.69 , -0.35) & -0.89 & (-1.41 , -0.39)\\

\toprule 
\multicolumn{7}{l}{\textbf{Leukemia}}\\
\toprule

Athey MC & 3.58 & NA & 2.69 & NA & 3.03 & (1.52 , 4.72)\\

Vanilla & 1.10 & (-2.25 , 3.22) & 0.19 & (-3.08 , 2.08) & 0.48 & (-1.90 , 1.90)\\

Space & 0.41 & (-7.32 , 2.77) & 0.39 & (-2.19 , 2.12) & 0.20 & (-2.38 , 1.80)\\

Space-Time 
 ICAR & -0.46 & (-2.97 , 1.98) & 0.56 & (-1.37 , 1.78) & 0.19 & (-1.45 , 1.53)\\

Space-Time 
 AR & -0.29 & (-1.16 , 0.49) & 0.16 & (-0.66 , 0.84) & -0.02 & (-0.62 , 0.50)\\

Space-Time 
 Lasso & -1.65 & (-4.54 , 0.50) & 1.32 & (0.28 , 2.13) & 0.16 & (-1.15 , 1.20)\\

Space-Time 
 Shrinkage & 1.73 & (0.87 , 2.66) & 0.11 & (-0.76 , 0.94) & 0.74 & (0.15 , 1.33)\\

\toprule
\end{tabular}

\end{table}

\subsection{Modeling the Observed Outcomes Under Treatment \label{s:modeled_Y1}}

To assess sensitivity of our ATT estimates to noise in the observed $Y_{it}=Y_{it}(1)$ values for treated units post-treatment, we conduct analyses that apply a priori spatio-temporal smoothing to these values. We perform this smoothing using a standard Bayesian spatio-temporal disease mapping model-- specifically the spatial autocorrelated first-order autoregressive process implemented in the \texttt{ST.CARar()} function in the \texttt{CARBayesST} R package \cite{JSSv084i09}. 

That is, we fit

$$
\mathrm{log}(E[Y_{it}(1)]) = \mathrm{X}_{it}'\beta + \psi_{it} + \mathrm{log}(\theta_{it})
$$
on the observed data for treated counties at post-treatment periods, where $\mathrm{X}_{it}$ is the percent of population that is Hispanic for county $i$ and time $t$, $\beta$ is its coefficient parameter, $\psi_{it}$ is a spatio-temporally autocorrelated random effect, and $\mathrm{log}(\theta_{it})$ represents the population offset. Using the \texttt{ST.CARar()} function, we specify a Poisson likelihood model with a log link function and 2,000 MCMC samples with 1,000 burn-in samples to be discarded.

\begin{figure}[H]
\centering
\subfloat[Lymphoma]
{\includegraphics[width=1\linewidth]{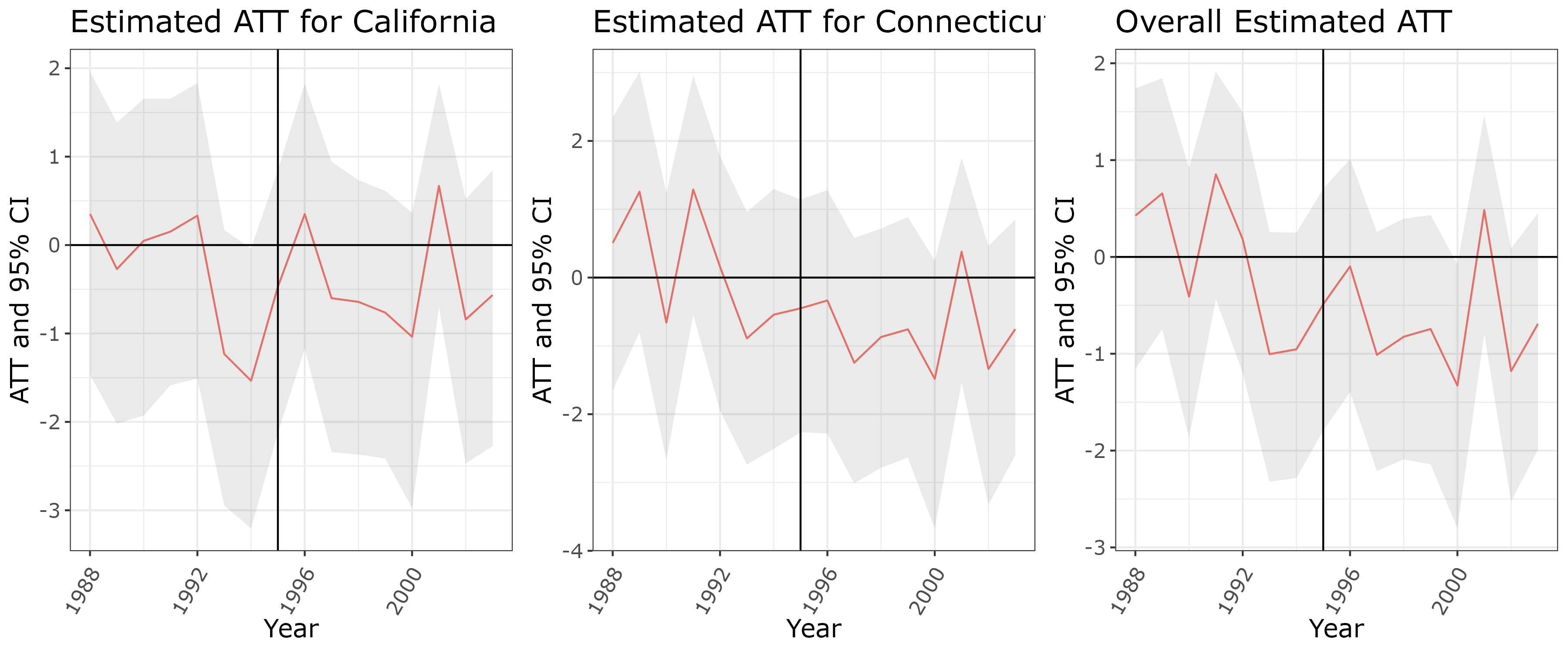}}

\subfloat[Leukemia]
{\includegraphics[width=1\linewidth]{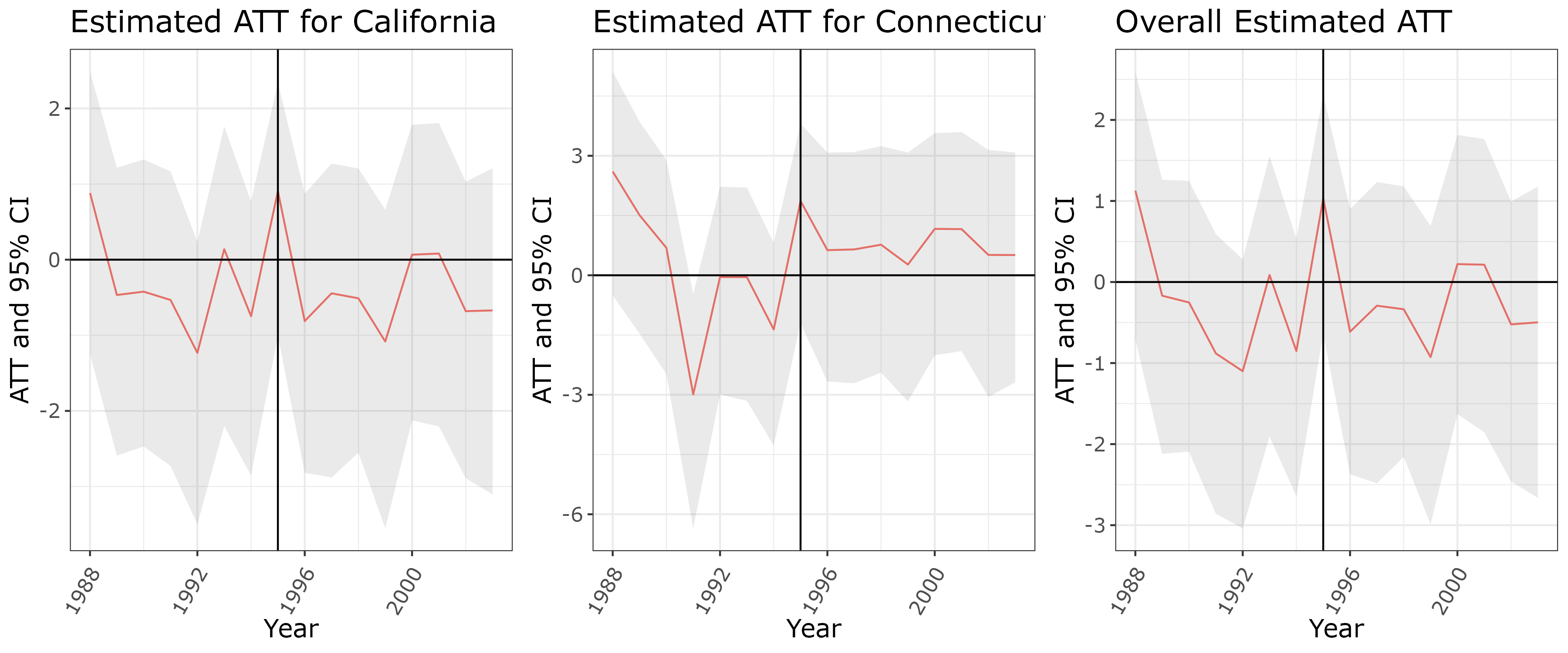}}

\caption[Estimated ATT Over Time]{Plots showing the estimated ATT over-time for incidence of lymphoma (a) and leukemia (b) when the observed outcomes under treatment are modeled. Space-Time AR Models were chosen based on evaluation of all models pre-treatment fits. \label{fig:ATT_time_modeled}}
\end{figure}

\begin{figure}[H]
  \centering
  \subfloat[Lymphoma]{\includegraphics[width=0.5\textwidth]{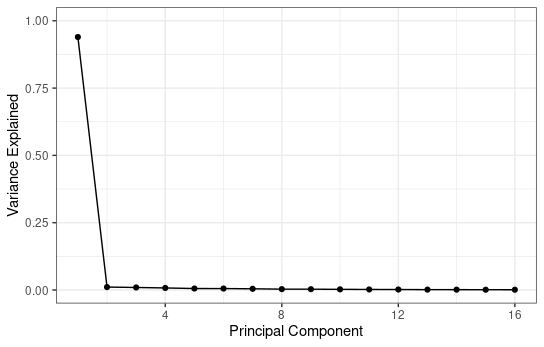}\label{fig:ScreeLymphoma}}
  \hfill
  \subfloat[Leukemia]{\includegraphics[width=0.5\textwidth]{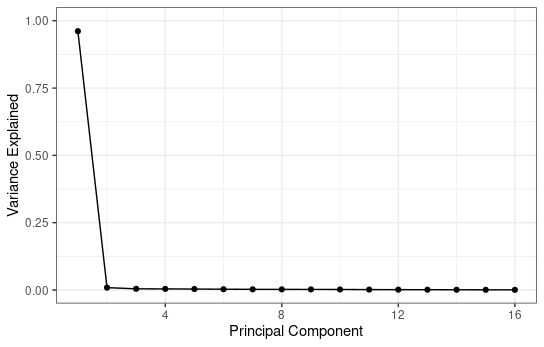}\label{fig:ScreeLeukemia}}
  \caption{Scree plots showing the percent of variance explained by each principal component for each panel data matrix: lymphoma (a) and leukemia (b).}
\end{figure}

Using the fitted values from this model ($\hat{Y}_{it}(1)$) in place of the observed outcomes under treatment, we calculate the ATT as

$$
\mathrm{ATT}_t^{(m)} = \frac{1}{N_\mathcal{W}}\sum_{i \in W}\Big[ \hat Y_{it}(1) - Y_{it}^{(m)}(0)\Big].
$$ Results are shown in Table \ref{tab:SuppATTLymphomaLeukemia_SensAnalysis_Y1}. Figure \ref{fig:ATT_time_modeled} shows the estimated ATT over time. This approach yields very similar results to the primary analysis presented in the main text (Section \ref{s:application}). 

\begin{table}[H]
\caption{Overall and state-specific estimated ATT results for each model applied to CYA lymphoma and leukemia incidence data without a temporal smoothing pre-processing step when the observed outcomes under treatment are modeled.}
\label{tab:SuppATTLymphomaLeukemia_SensAnalysis_Y1}
\begin{tabular}{l c c c c c c}
\toprule
\multirow{2}{*}{ Model }  & \multicolumn{2}{c}{ California } & \multicolumn{2}{c}{ Connecticut } & \multicolumn{2}{c}{ Overall }\\

& ATT & 95\% CI & ATT & 95\% CI & ATT & 95\% CI\\
\toprule 
\multicolumn{7}{l}{\textbf{Lymphoma}}\\
\toprule 
Athey MC & -0.15 & NA & -1.26 & NA & -0.84 & (-2.22 , -0.10)\\

Vanilla & -0.63 & (-1.80 , 0.27) & -1.47 & (-2.61 , -0.47) & -1.14 & (-2.09 , -0.37)\\

Space & -0.80 & (-2.52 , 0.35) & -1.43 & (-2.70 , -0.35) & -1.21 & (-2.30 , -0.36)\\

Space-Time 
 ICAR & -0.52 & (-1.66 , 0.43) & -1.37 & (-2.60 , -0.36) & -1.05 & (-1.96 , -0.30)\\

Space-Time 
 AR & -0.49 & (-0.98 , 0.04) & -0.79 & (-1.40 , -0.21) & -0.67 & (-1.08 , -0.24)\\

Space-Time 
 Lasso & -0.65 & (-1.57 , 0.17) & -1.40 & (-2.29 , -0.45) & -1.12 & (-1.82 , -0.39)\\

Space-Time 
 Shrinkage & -0.33 & (-1.04 , 0.40) & -1.12 & (-1.94 , -0.28) & -0.83 & (-1.37 , -0.24)\\

\toprule 
\multicolumn{7}{l}{\textbf{Leukemia}}\\
\toprule
Athey MC & 0.09 & NA & 3.73 & NA & 0.63 & (-0.74 , 2.57)\\

Vanilla & 0.19 & (-1.19 , 1.28) & 0.86 & (-1.00 , 2.32) & 0.29 & (-1.00 , 1.25)\\

Space & 0.57 & (-0.66 , 1.35) & 0.72 & (-0.73 , 1.86) & 0.57 & (-0.52 , 1.28)\\

Space-Time 
 ICAR & 0.49 & (-0.66 , 1.44) & 0.90 & (-0.33 , 1.95) & 0.55 & (-0.46 , 1.40)\\

Space-Time 
 AR & -0.40 & (-1.15 , 0.22) & 0.73 & (-0.23 , 1.65) & -0.23 & (-0.90 , 0.35)\\

Space-Time 
 Lasso &  0.33 & (-0.77 , 1.22) & 0.91 & (-0.91 , 2.22) & 0.39 & (-0.64 , 1.25)\\

Space-Time 
 Shrinkage & -0.01 & (-1.26 , 1.20) & 0.87 & (-0.54 , 2.23) & 0.09 & (-1.07 , 1.28)\\

\toprule
\end{tabular}
\end{table}

\newpage
\subsection{Additional Tables and Figures}

\begin{table}[H]
\caption{Counties included in the federal RFG program in 1995. }
\label{tab:FedRFG}
\center
\begin{tabular}{ p{3cm} p{5.5cm} }

 \toprule
 State& Counties\\
  \toprule
     \multirow{3}{5em}{California} & Los Angeles, Orange,\\ & Ventura,  San Bernardino, \\
     & Riverside, San Diego\\

\hline
 \multirow{4}{5em}{Connecticut} & Hartford, Middlesex,  \\ 
    & New Haven, New London,\\
    & Tolland, Windham, \\
    & Fairfield, Litchfield \\

 \hline
 \multirow{8}{5em}{New Jersey} & Bergen, Essex,  \\ 
    & Hudson, Hunterdon, \\
    & Middlesex, Monmouth,\\
    & Morris, Ocean, Passaic \\
    & Somerset, Sussex, Union \\
    & Burlington, Camden, \\
    & Cumberland, Gloucester, \\
    & Mercer, Salem\\
 \hline
 
\multirow{5}{5em}{New York} & Bronx, Kings, Nassau, \\ 
     & New York, Orange, Putnam,\\
     & Queens, Richmond, \\
     & Rockland, Suffolk,\\
     & Westchester\\
 \hline
 
\multirow{1}{5em}{Delaware} & Kent, New Castle \\ 
\hline

\multirow{3}{5em}{Maryland} & Cecil Anne Arundel,\\
& Baltimore, Carroll, \\
& Harford, Howard\\
\hline

\multirow{2}{5em}{Pennsylvania} & Bucks, Chester, Delaware\\
& Montgomery, Philadelphia\\
\hline

\multirow{1}{5em}{Indiana} & Lake, Porter\\
\hline

\multirow{3}{5em}{Texas} & Brazoria, Chambers,\\
& Fort Bend, Galveston, Harris,\\
& Liberty, Montgomery, Waller\\
\hline

\multirow{3}{5em}{Wisconsin} & Kenosha, Milwaukee,\\
& Ozaukee, Racine, \\
& Washington, Waukesha\\
 \toprule

\end{tabular}

\end{table}

\newpage

\begin{center}

\begin{longtable}{ 
p{4.6cm}cC{2.5cm}C{2.2cm}C{2.5cm}C{2.3cm} }

\caption[]{Simulation results: Average absolute percent bias of the counterfactual estimates from each method applied with different specifications of the number of factors, $K$ (rows), in simulated data emulating rare and non-rare outcomes, with and without a temporal smoothing pre-processing step (columns), with respective 25$^{\mathrm{th}}$ and 75$^{\mathrm{th}}$ percentiles in  parentheses.} \label{tab:iqr} \\

\toprule
\multirow{2}{*}{Model} & \multirow{2}{*}{ K } & Non-Rare \&  Non-Smoothed  & Non-Rare \& Smoothed & Rare \& Non-Smoothed & Rare \& Smoothed\\
\toprule
\endfirsthead
\toprule
\multirow{2}{*}{Model} & \multirow{2}{*}{ K } & Non-Rare \&  Non-Smoothed  & Non-Rare \& Smoothed & Rare \& Non-Smoothed & Rare \& Smoothed\\
\toprule
\endhead

 \multicolumn{6}{r}{{Continued on next page}} \\ 
\endfoot

\endlastfoot

\multicolumn{6}{l}{\textbf{Bayesian MC Models}} \\ \bottomrule
\multirow{6}{6em}{\hspace{.4cm}1. Vanilla} & \multirow{2}{.5em} 1 & 11.56  & 11.11  & 29.65  & 29.87 \\
& & (8.74, 14.22) & (8.65,13.72) &  (21.51,38.80) & (22.24,38.38)\\
 & \multirow{2}{.5em}3 & \textbf{10.50} & 11.33 & 32.77 & 30.90\\
& & (7.57,13.57) & (7.99,14.39) & (22.99,40.72) & (22.44,39.8) \\
 & \multirow{2}{.5em} 7 & 11.12 & 11.64 & 33.89 & 34.18 \\
 & & (8.85,14.51) & (8.59,14.76) & (26.16,44.98) & (24.33,39.98) \\
\hline
\multirow{6}{6em}{\hspace{.4cm}2. Space} & \multirow{2}{.5em} 1 & \textbf{10.29} & \textbf{10.50} & 29.60 & \textbf{28.51} \\
& & (7.40,13.42) & (7.90,13.61) & (22.61,36.52) & (21.73,37.52) \\

 & \multirow{2}{.5em} 3 & \textbf{10.69} & \textbf{10.45}  & \textbf{29.38}  & \textbf{28.33} \\
 & & (7.58,13.60) & (7.48,13.11) & (22.06,37.44) & (20.40,36.40) \\

 & \multirow{2}{.5em} 7 & 10.79  & \textbf{10.53}  & \textbf{28.62} & 28.96\\
 & & (7.53,13.56) & (7.93,13.39) & (22.53,38.10) & (21.23,38.09) \\
\hline
\multirow{6}{*}{\hspace{.4cm}3. Space-Time ICAR} & \multirow{2}{.5em} 1 & 10.90 & 10.69 & 30.42 & \textbf{28.04}\\
& & (7.92,13.42) & (7.14,13.71) & (22.23,37.79) & (20.68,35.00) \\

 & \multirow{2}{.5em} 3 & 10.83 & 10.66 & 29.45 & 29.89\\
& & (8.05,13.19) & (7.75,13.04) & (21.1,38.21) & (20.79,38.10) \\
 & \multirow{2}{.5em} 7 & 10.76 & 10.56 & \textbf{29.18} & 29.33 \\

 & & (8.46,13.50) & (7.64,13.45) & (20.89,38.37) & (21.21,38.24) \\
\hline
\multirow{6}{*}{\hspace{.4cm}4. Space-Time AR} & \multirow{2}{.5em} 1 & 10.87 & \textbf{10.46} & 30.46 & 29.94\\
& & (7.37,13.98) & (7.79,13.27) & (21.22,38.74) & (22.41,36.4) \\

 & \multirow{2}{.5em} 3 & \textbf{10.32} & 10.56 & \textbf{28.94} & \textbf{28.87}\\
 & & (7.57,13.43) & (7.66,13.92) & (22.24,37.55) & (22.93,36.94) \\

 & \multirow{2}{.5em} 7 & 10.95 & \textbf{10.29} & \textbf{28.99} & \textbf{28.00}\\
 & & (8.09,13.27) & (7.35,13.77) & (21.23,36.76) & (23.37,35.60) \\
\hline
\multirow{6}{*}{\hspace{.4cm}5. Space-Time Lasso} & \multirow{2}{.5em} 1 & 13.35 & 12.32 & 31.02 & 30.01 \\
& & (8.84,17.06) & (8.46,14.88) & (23.34,41.84) & (22.80,41.91) \\

 & \multirow{2}{.5em} 3 & 12.59 & 13.19 & 37.86 & 37.26 \\
 & & (8.88,17.28) & (9.21,18.94) & (28.45,45.20) & (29.16,47.34) \\

 & \multirow{2}{.5em} 7 & 16.30  & 16.47 & 57.03 & 46.27 \\
& & (11.63,24.55) & (10.93,20.08) & (42.77,76.29) & (32.51,67.58) \\
\hline
\multirow{2}{*}{\hspace{.4cm}6. Space-Time Shrinkage} & \multirow{2}{.5em} 7 & \textbf{10.66} & 10.65 & 29.94 & 29.38 \\
& & (7.48,14.03) & (7.47,13.75) & (22.03,36.77) & (20.28,36.06) \\
\toprule 
\newpage

\multicolumn{6}{l}{\textbf{Existing Methods}} \\* \bottomrule
\multirow{6}{6em}{\hspace{.4cm}Athey MC} & \multirow{2}{.5em} 1 & 13.78 & 13.89 & 32.34 & 34.98 \\*
& & (10.77,18.12) & (9.83,18.64) & (25.73,41.52) & (24.74,40.49) \\*

& \multirow{2}{.5em} 3 & 14.14 & 14.43 & 32.22 & 34.70 \\*
& & (11.11,18.05) & (10.19,18.12) & (24.92,41.15) & (23.55,39.96) \\*

& \multirow{2}{.5em} 7 & 14.75  & 14.49 & 31.73  & 33.89 \\*
& & (11.12,18.12) & (10.68,18.63) & (24.64,40.20) & (24.94,39.40) \\
\hline
\multirow{2}{6em}{\hspace{.4cm}GSC} & \multirow{2}{1.5em} {NA} & 14.11 & 14.13  & 31.92  & 34.70 \\
& & (11.08,18.39) & (9.95,18.47) & (24.97,42.62) & (25.52,42.19) \\
\toprule
\multirow{2}{6em}{\hspace{.4cm}ASC} & \multirow{2}{1.5em} {NA} & 14.04  & 13.13 & 36.14  & 33.28 \\*
& & (10.63,17.88) & (9.07,16.75) & (27.74,49.70) & (24.20,42.36) \\
\hline
\multirow{6}{6em}{\hspace{.4cm}ALS MC} & \multirow{2}{.5em} 1 & 12.14  & 12.48 & 34.25 & 33.01 \\
& & (9.97,15.98) & (9.68,16.19) & (25.13,44.00) & (24.61,44.17) \\

& \multirow{2}{.5em} 3 & 14.80  & 13.78  & 37.97  & 35.06 \\
& & (10.39,20.82) & (10.24,18.69) & (27.07,51.53) & (26.67,45.79) \\

 & \multirow{2}{.5em} 7 & 21.90  & 19.92  & 52.46  & 42.89 \\
 & & (13.31,30.35) & (13.67,28.28) & (41.52,71.03) & (26.93,55.54) \\
\hline
\multirow{2}{*}{\hspace{.4cm}Soft Impute MC} & \multirow{2}{1.5em} {NA} & 99.33  & 99.33  & 96.14  & 96.21 \\
& & (99.28,99.37) & (99.28,99.38) & (95.53,96.91) & (95.59,96.87) \\
\hline
\multirow{2}{*}{\hspace{.4cm}Nuclear Norm MC} & \multirow{2}{1.5em} {NA} & 12.99 & 12.89 & 34.58  & 37.06\\
& & (10.36,17.42) & (9.34,18.37) & (27.78,45.03) & (24.56,42.25) \\
\hline
\multirow{2}{6em}{\hspace{.4cm}SVT MC} & \multirow{2}{1.5em} {NA} & 36.35 & 36.24 & 41.68 & 40.26\\
& & (30.98,41.66) & (31.11,41.41) & (28.84,53.32) & (28.75,54.15) \\
\hline
\end{longtable}
\end{center}

\begin{table}[H]
\caption{Simulation results: Average absolute percent bias of the counterfactual estimates from each method applied with different specifications of the number of factors, $K$ (rows), in simulated data emulating rare and non-rare outcomes, with and without a temporal smoothing pre-processing step (columns), for $\tau^2=0.08$ in the simulation structure.}
\label{tab:absPercBiasY0_alttau}
\begin{center}
\begin{tabular}{ 
p{4.6cm}cC{2.5cm}C{2.2cm}C{2.5cm}C{2cm} } 
\toprule
\multirow{2}{*}{Model} & \multirow{2}{*}{ K } & Non-Rare \&  Non-Smoothed  & Non-Rare \& Smoothed & Rare \& Non-Smoothed & Rare \& Smoothed\\
\toprule
\multicolumn{6}{l}{\textbf{Bayesian MC Models}} \\ \bottomrule
\multirow{3}{6em}{\hspace{.4cm}1. Vanilla} & 1 & 9.78 & 9.47 & 26.88 & 26.09\\
 & 3 & \textbf{8.90} & \textbf{9.13} & 26.85 & 26.30\\
 & 7 & 9.45 & 9.90 & 27.48 & 28.53\\
\hline
\multirow{3}{6em}{\hspace{.4cm}2. Space} & 1 & 9.54 & 9.56 & 25.71 & \textbf{25.59}\\
 & 3 & 9.38 & \textbf{9.06} & 25.73 & 26.06\\
 & 7 & \textbf{9.08} & \textbf{9.31} & 26.14 & \textbf{25.87}\\
\hline
\multirow{3}{*}{\hspace{.4cm}3. Space-Time ICAR} & 1 & 9.19 & 9.47 & \textbf{24.17} & 26.00\\
& 3 & \textbf{9.09} & 9.46 & \textbf{24.17} & 26.07\\
 & 7 & 9.18 & \textbf{8.98} & \textbf{25.29} & \textbf{25.69}\\
\hline
\multirow{3}{*}{\hspace{.4cm}4. Space-Time AR} & 1 & 9.22 & 9.69 & 25.71 & \textbf{25.96}\\
 & 3 & \textbf{9.04} & \textbf{9.17} & \textbf{25.41} & 25.98\\
 & 7 & \textbf{8.83} & 9.74 & 25.57 & 26.00\\
\hline
\multirow{3}{*}{\hspace{.4cm}5. Space-Time Lasso} & 1 & 10.34 & 10.81 & 28.68 & 27.75\\
 & 3 & 12.76 & 12.04 & 34.90 & 28.83\\
 & 7 & 14.44 & 14.26 & 49.48 & 44.62\\
\hline
\hspace{.4cm}6. Space-Time Shrinkage  & 7 & 9.52 & 9.58 & \textbf{25.25} & \textbf{25.69}\\
\toprule
\multicolumn{6}{l}{\textbf{Existing Methods}} \\ \bottomrule
\multirow{3}{6em}{\hspace{.5cm}Athey MC} & 1 & 11.36 & 11.09 & 28.66 & 29.45\\
 & 3 & 11.26 & 11.14 & 28.45 & 28.49\\
 & 7 & 11.18 & 11.56 & 28.66 & 29.02\\
\hline
\multirow{1}{6em}{\hspace{.5cm}GSC} & NA & 11.66 & 12.57 & 31.09 & 29.47\\
\hline
\multirow{1}{6em}{\hspace{.5cm}ASC} & NA & 12.22 & 13.15 & 35.94 & 28.49\\
\hline
\multirow{3}{6em}{\hspace{.5cm}ALS MC} & 1 & 11.26 & 11.15 & 29.17 & 29.58\\
 & 3 & 14.35 & 12.28 & 36.28 & 32.79\\
 & 7 & 19.47 & 17.44 & 44.62 & 36.13\\
\hline
\multirow{1}{*}{\hspace{.3cm}Soft Impute MC} & NA & 99.43 & 99.43 & 96.80 & 96.73\\
\hline
\multirow{1}{*}{\hspace{.3cm}Nuclear Norm MC} & NA & 13.11 & 11.75 & 32.62 & 31.57\\
\hline
\multirow{1}{6em}{\hspace{.5cm}SVT MC} & NA & 28.86 & 28.71 & 27.91 & 28.25\\
\hline
\end{tabular}
\end{center}
\end{table}

\begin{table}[H]
\caption{Average R-hat for each model's counterfactual predictions applied with different specifications of the number of factors, $K$ (rows), in CYA lymphoma and leukemia incidence data, with and without a temporal smoothing pre-processing step (columns).\label{tab:rhat}}
\begin{center}

\begin{tabular}{ p{4cm}cC{2.5cm}C{2.5cm}C{2.5cm}C{2cm} } 
\toprule
\multirow{2}{*}{Model} & \multirow{2}{*}{ K } & Lymphoma \&  Non-Smoothed  & Lymphoma \& Smoothed & Leukemia \& Non-Smoothed & Leukemia \& Smoothed\\
\toprule
\multirow{2}{6em}{Vanilla} & 2 & 1.00 & 1.00 & 1.00 & 1.00\\
 & 3 & 1.00 & 1.00 & 1.00 & 1.00\\
\hline
\multirow{2}{6em}{Space} & 2 & 1.00 & 1.01 & 1.01 & 1.01\\
 & 3 & 1.01 & 1.01 & 1.01 & 1.01\\
\hline
\multirow{2}{*}{Space-Time ICAR} & 2 & 1.01 & 1.03 & 1.01 & 1.01\\
 & 3 & 1.01 & 1.02 & 1.01 & 1.02\\
\hline
\multirow{2}{*}{Space-Time AR} & 2 & 1.00 & 1.00 & 1.00 & 1.00\\
  & 3 & 1.00 & 1.00 & 1.00 & 1.00\\
\hline
\multirow{2}{*}{Space-Time Lasso} & 2 & 1.02 & 1.02 & 1.02 & 1.02\\
 & 3 & 1.04 & 1.03 & 1.05 & 1.02\\
\hline
\multirow{2}{*}{Space-Time Shrinkage} & 3 & 1.02 & 1.02 & 1.01 & 1.01\\
 & 4 & 1.01 & 1.02 & 1.02 & 1.02\\

\toprule
\end{tabular}

\end{center}

\end{table}

\begin{table}[H]
\caption{ Overall and state-specific estimated ATT results for each model applied with different specifications of the number of factors, $K$ (rows), in CYA lymphoma and leukemia incidence data with a temporal smoothing pre-processing step.}
\label{tab:ATTLymphomaLeukemiaSmooth}
\begin{tabular}{l c c c c c c c}
\toprule
\multirow{2}{*}{ Model } & \multirow{2}{*}{ K } & \multicolumn{2}{c}{ California } & \multicolumn{2}{c}{ Connecticut } & \multicolumn{2}{c}{ Overall }\\

& & ATT & 95\% CI & ATT & 95\% CI & ATT & 95\% CI\\
\toprule 
\multicolumn{7}{l}{\textbf{Lymphoma}}\\
\toprule 
\multirow{2}{6em}{Athey MC} & 2 & -0.37 & NA & -1.42 & NA & -1.02 & (-2.08 , 0.01)\\
 & 3 & -0.37 & NA & -1.42 & NA & -1.02 & (-2.07 , 0.00)\\
\hline

\multirow{2}{6em}{Vanilla} & 2 & -0.86 & (-2.59 , 0.43) & -1.96 & (-4.06 , -0.62) & -1.55 & (-3.09 , -0.52)\\
& 3 & -0.96 & (-3.04 , 0.36) & -1.94 & (-4.18 , -0.59) & -1.59 & (-3.26 , -0.48)\\
\hline
\multirow{2}{6em}{Space} & 2 & -0.96 & (-2.49 , 0.14) & -1.68 & (-3.06 , -0.54) & -1.39 & (-2.57 , -0.46)\\
 & 3 & -1.43 & (-42.55 , 0.26) & -1.70 & (-3.48 , -0.45) & -1.71 & (-17.5 , -0.30)\\
\hline
\multirow{2}{*}{Space-Time ICAR} & 2 & -1.42 & (-4.25 , 0.23) & -1.56 & (-2.50 , -0.67) & -1.51 & (-2.95 , -0.66)\\
 & 3 & -1.83 & (-6.44 , 0.37) & -1.81 & (-3.78 , -0.26) & -1.77 & (-4.34 , -0.21)\\
\hline
\multirow{2}{*}{Space-Time AR} & 2 & -1.13 & (-1.69 , -0.61) & -1.18 & (-1.76 , -0.60) & -1.15 & (-1.59 , -0.74)\\
 & 3 & -1.00 & (-1.46 , -0.53) & -2.80 & (-3.49 , -2.13) & -2.12 & (-2.58 , -1.67)\\
\hline
\multirow{2}{*}{Space-Time Lasso} & 2 & -1.93  & (-3.10 , -0.95) & -1.74 & (-2.79 , -0.79) & -1.84 & (-2.61 , -1.09)\\
 & 3 & -0.85 & (-1.56 , -0.02) & -5.45 & (-13.87 , -2.38) & -3.65 & (-8.64 , -1.86)\\
\hline
\multirow{2}{*}{Space-Time Shrinkage} & 3 & -0.83 & (-1.43 , -0.28) & -1.32 & (-2.32 , -0.54) & -1.15 & (-1.80 , -0.57)\\
 & 4 & -0.75 & (-1.67 , -0.17) & -1.69 & (-2.46 , -0.98) & -1.37 & (-2.03 , -0.82)\\
\toprule 
\multicolumn{7}{l}{\textbf{Leukemia}}\\
\toprule 

\multirow{2}{6em}{Athey MC} & 2 & 0.95 & NA & 3.85 & NA & 1.38 & (-0.47 , 2.70)\\
 & 3 & 0.95 & NA & 3.85 & NA & 1.38 & (-0.48 , 2.71)\\
\hline

\multirow{2}{6em}{Vanilla} & 2 & 0.78 & (-0.51 , 1.81) & 1.27 & (-0.88 , 2.64) & 0.84 & (-0.25 , 1.78)\\
& 3 & 0.69 & (-0.61 , 1.78) & 1.21 & (-1.04 , 2.64) & 0.77 & (-0.51 , 1.73)\\
\hline
\multirow{2}{6em}{Space} & 2 & 1.09 & (0.12 , 1.81) & 1.05 & (-0.69 , 2.41) & 1.09 & (0.18 , 1.72)\\
 & 3 & 1.01 & (0.02 , 1.84) & 1.26 & (-0.13 , 2.48) & 1.05 & (0.19 , 1.80)\\
\hline
\multirow{2}{*}{Space-Time ICAR} & 2 & 1.12 & (0.16 , 1.92) & 1.72 & (0.27 , 2.86) & 1.23 & (0.29 , 1.93)\\
 & 3 & 0.96 & (0.01 , 1.73) & 1.51 & (0.27 , 2.58) & 1.03 & (0.22 , 1.75)\\
\hline
\multirow{2}{*}{Space-Time AR} & 2 & 0.96 & (0.41 , 1.43) & 2.63 & (1.91 , 3.23) & 1.21 & (0.74 , 1.61)\\
& 3 & 1.49 & (0.96 , 1.97) & 1.68 & (0.81 , 2.47) & 1.51 & (1.04 , 1.96)\\
\hline
\multirow{2}{*}{Space-Time Lasso} & 2 & 0.25  & (-0.97 , 1.26) & -3.42 & (-20.87 , 0.62) & -0.37 & (-3.31 , 0.95)\\
 & 3 & 0.89 & (-5.36 , 1.43) & 1.02 & (-0.56 , 2.25) & 0.11 & (-4.40 , 1.47)\\
\hline
\multirow{2}{*}{Space-Time Shrinkage} & 3 & 0.84 & (-0.22 , 1.74) & 1.61 & (0.16 , 2.76) & 0.94 & (-0.01 , 1.76)\\
 & 4 &  -0.01  & (0.08 , 1.76) & 1.46 & (0.18 , 2.54) & 0.96 & (0.25 , 1.83)\\

\toprule
\end{tabular}

\end{table}

\begin{table}[H]
\caption{ Overall and state-specific estimated ATT results for each model applied with different specifications of the number of factors, $K$ (rows), in CYA lymphoma and leukemia incidence data without a temporal smoothing pre-processing step.}
\label{tab:ATT_SensAnalysis_K}
\begin{tabular}{l c c c c c c c}
\toprule
\multirow{2}{*}{ Model } & \multirow{2}{*}{ K } & \multicolumn{2}{c}{ California } & \multicolumn{2}{c}{ Connecticut } & \multicolumn{2}{c}{ Overall }\\

& & ATT & 95\% CI & ATT & 95\% CI & ATT & 95\% CI\\
\toprule 
\multicolumn{7}{l}{\textbf{Lymphoma}}\\
\toprule 

Athey MC & 3 & -0.46 & NA & -1.60 & NA & -1.16 & (-2.21 , -0.12)\\

Vanilla & 3 & -1.05 & (-2.34 , 0.00) & -1.88 & (-3.25 , -0.83) & -1.59 & (-2.61 , -0.74)\\

Space & 3 & -1.24 & (-3.40 , 0.49) & -1.85 & (-3.77 , -0.73) & -1.65 & (-2.96 , -0.60)\\

Space-Time 
 ICAR & 3 & -1.00 & (-2.79 , 0.47) & -1.71 & (-2.72 , -0.78) & -1.47 & (-2.39 , -0.50)\\

Space-Time 
 AR & 3 & -0.72 & (-1.21 , -0.20) & -0.89 & (-1.47 , -0.38) & -0.83 & (-1.23 , -0.45)\\

Space-Time 
 Lasso & 3 & -0.93 & (-1.79 , -0.14) & -1.72 & (-2.75 , -0.65) & -1.40 & (-2.22 , -0.67)\\

Space-Time 
 Shrinkage  & 4 & -0.82 & (-1.66 , -0.08) & -1.99 & (-2.87 , -1.26) & -1.55 & (-2.18 , -0.98)\\

\toprule 
\multicolumn{7}{l}{\textbf{Leukemia}}\\
\toprule 
Athey MC & 3 & 0.36 & NA & 4.14 & NA & 0.92 & (-0.69 , 2.52)\\

Vanilla & 3 & 0.50 & (-0.88 , 1.62) & 1.20 & (-0.94 , 2.64) & 0.60 & (-0.65 , 1.62)\\

Space & 3 & 0.74 & (-0.41 , 1.60) & 0.85 & (-1.26 , 2.28) & 0.76 & (-0.35 , 1.61)\\

Space-Time 
 ICAR & 3 & 0.88 & (-0.03 , 1.73) & 1.53 & (0.19 , 2.54) & 0.97 & (0.18 , 1.72)\\

Space-Time 
 AR & 3 & 0.94 & (0.39 , 1.45) & 1.82 & (0.95 , 2.64) & 1.07 & (0.58 , 1.52)\\

Space-Time 
 Lasso & 3  & -13.11 & (-179.29 , -1.22) & -1.45 & (-8.17 , 1.97) & -11.39 & (-153.18 , -0.80)\\

Space-Time 
 Shrinkage  & 4 &  0.45 & (-0.62 , 1.30) & 1.66 & (0.34 , 2.70) & 0.63 & (-0.30 , 1.38)\\

\toprule
\end{tabular}

\end{table}

\end{document}